\documentclass[epsfig,12pt]{article}

\usepackage[dvips]{graphicx}
\usepackage{subfig}
\begin{document}

\centerline{ \Large \bf  Classical Particle in a Box with Random
Potential:} \centerline{\bf \Large  exploiting rotational symmetry
of replicated Hamiltonian.}

\vskip 0.4cm

\centerline{\large \bf Yan V. Fyodorov$^{a,b}$ and H.-J.
Sommers$^{c}$}

\vskip 0.3cm

\centerline{$^a$ Institut f\"{u}r Theoretische Physik,
Universit\"{a}t zu K\"{o}ln, D-50937 K\"{o}ln, Germany}

\centerline{$^b$ School of Mathematical Sciences, University of
Nottingham, Nottingham NG72RD, England\footnote{permanent address}}

\centerline{$^c$ Fachbereich Physik,  Universit\"{a}t
Duisburg-Essen, D-47048 Duisburg, Germany}

\vskip 0.3cm

\begin{abstract}
We provide a detailed discussion of the replica approach to
thermodynamics of a single classical particle placed in a random
Gaussian $N\,(>>1)-$dimensional potential inside a spherical box
of a finite radius $L=R\sqrt{N}$. Earlier solutions of $R=\infty$
version of this model were based on applying the Gaussian
Variational Ansatz (GVA) to the replicated partition function, and
revealed a possibility of glassy phases at low temperatures. For a
general $R$, we show how to utilize instead the underlying
rotational symmetry and to arrive to a compact expression for the
free energy in the limit $N\to\infty$ directly, without any need
for intermediate variational approximations. This method reveals a
striking similarity with the much-studied spherical model of spin
glasses. Depending on the competition between the radius $R$ and
the curvature of the parabolic confining potential $\mu\ge 0$, as
well as on the three types of disorder - short-ranged,
long-ranged, and logarithmic - the phase diagram of the system in
the $(\mu,T)$ plane undergoes considerable modifications. In the
limit of infinite confinement radius our analysis confirms all
previous results obtained by GVA. The paper has also a
considerable pedagogical component by providing an extended
presentation of technical details which are not always easy to
find in the existing literature.

\end{abstract}

\section{Introduction}

In this paper we perform a detailed study of thermodynamics of a
single classical particle confined to a spherical box filled in
with an energy landscape described by a random Gaussian function
${\cal H}$ of $N$ real variables ${\bf x}=(x_1,...,x_N)$. Although
the problem is meaningful for any $N$, we eventually will be
mainly concerned with the limit of large $N\gg 1$ where we will be
able to develop a systematic method of analysis. In fact, it is
well known that such simple yet non-trivial models play a role of
a laboratory for developing the methods allowing one to deal with
problems of statistical mechanics where an interplay between
thermal fluctuations and those due to quenched disorder is
essential. The paradigmatic example of systems of this sort are
spin glasses \cite{spinglass}, but similar effects are frequently
operational for polymers' behaviour in random environment, for
phase separating interfaces in random field models or for elastic
manifolds pinned by random impurities. In general, presence of
quenched disorder leaves no choice but to employ the so-called
replica trick, which is a heuristic way of extracting the averaged
free energy of the system from moments of the partition function.
The book \cite{Dotsenko} may serve as a modern introduction to
this problematic.

Both conceptual and technical difficulties of dealing with
statistical mechanics of disordered systems stem from the fact
that many features of their low temperature dynamics and
thermodynamics are dominated by presence of a huge number of
metastable states (both minima and saddle points of various types
) in the energy functional in configuration space. At finite
temperature those features may generate a complicated free energy
landscape, and that structure is responsible both for unusual
equilibrium properties (an "ergodicity breaking", see
\cite{spinglass,Dotsenko}) as well as for a complicated long-time
dynamical behaviour. The latter manifests itself, in particular,
through slow relaxation and aging effects \cite{longtime}.  At the
level of static properties the broken ergodicity is reflected in
an intricate pattern of spontaneous replica symmetry breaking
discovered originally by Parisi \cite{Parisi} in the framework of
the Sherrington-Kirkpatrick \cite{SK} model of spin glass with
infinite-range interactions. Attempts to relate that picture to
the properties of stationary points of the corresponding free
energy landscape started long ago from the pioneering TAP
paper\cite{TAP} and is still a rather active field of research,
see e.g. \cite{metast1,metast2} and references therein for a
general account, and \cite{fluc,BrayDean,FSW} for a treatment of
the model of the present type.

In this broad context, the model of a single particle in a random
potential has played an important role in physics of disordered
system.  It has enjoyed quite a long history of research starting
from early works by Mezard and Parisi \cite{MP}, and Engel
\cite{Engel} on static properties of such a system, followed by
Franz and Mezard \cite{FM} and Cugliandolo and Le Doussal
\cite{longtime} papers on the corresponding dynamics. To have bona
fide thermodynamics one has to ensure that the partition function
$Z$ of the model is well-defined for any realization of the random
potential. This is usually achieved by introducing a sort of a
confining potential $V_{con}({\bf x})$ which prohibits escape of the
particle to infinity. The standard choice is to use a parabolic
potential uniform in all directions $V_{con}({\bf
x})=\frac{\mu}{2}{\bf x}^2$. The curvature $\mu>0$ then plays
together with the temperature $T$ a role of the main control
parameter of the system, and one of standard goals of the theory is
to investigate the phase diagram in the $(\mu,T)$ plane.
Technically, the problem amounts to calculating the ensemble average
of the equilibrium free energy $F=-T\ln{\,Z\,}$.  The replica trick
allows one to represent integer moments $\langle Z^n\rangle$ of the
partition function in a form of some multivariable non-Gaussian
integrals, and one then faces the usual problem of finding ways of
evaluating those integrals for need of performing the replica limit
$n\to 0$. The crucial step allowing one to achieve further progress
was suggested by Mezard and Parisi \cite{MP} and is widely known as
the Gaussian Variational Ansatz (GVA). Loosely speaking, it amounts
to replacing the non-Gaussian integrands with trial Gaussian ones
and employing the Feynman-Bogoliubov  variational procedure to find
best possible Gaussian approximation to the true free energy in the
replica limit, see \cite{MP,Engel} for a detailed discussion.
Proceeding in this way, it turned out to be possible to employ again
the Parisi scheme of spontaneous replica symmetry breaking at the
level of the trial free energy, and to reveal the glassy nature of
the lower-temperature phase of the model. Since its induction GVA
method became one of the most popular tools of dealing with quenched
disorder. It allowed not only to get useful insights into static
properties of glassy systems of quite a diverse nature (see e.g.
\cite{multif,loc,korsh,giam,garel}), but was eventually adopted to
study the corresponding nonequilibrium dynamics as well
\cite{longtime,dean,chain}. Such a flexibility of GVA is based on
the fact that it can be formally applied in any dimension $N$. The
accumulated experience shows that the physical quantities calculated
in such a way usually agree well with the results of numerical
simulations on qualitative, and sometimes even quantitative level.
As a result, GVA presently serves as a standard reference point for
comparison with any new approach to the problem, as e.g. with a
recently developed functional renormalization group method
\cite{frg}.

One should nevertheless recall that the exploitation of GVA lacks
formal mathematical justification for finite dimensions. To this
end, Mezard and Parisi provided an argument\cite{MP}, see also
\cite{Goldsch}, that being a variant of the Hartree-Fock approach
the method should actually yield exact results in the limit $N\to
\infty$.
 Another intrinsic
feature of GVA is that the whole procedure is essentially based on
a possibility to evaluate explicitly some intermediate Gaussian
integrals. For the models defined in a restricted geometry
presence of a geometric confinement may make applications of GVA
less convenient. This is precisely the case for the class of
models to be treated in the present paper: a particle confined to
an impenetrable spherical box of some finite radius
$L=R\sqrt{N},\,\, R<\infty$ filled with a Gaussian random
potential. To make contact with previous works on the problem we
retain also the parabolic confinement term, so that our model is
characterised by both $R$ and $\mu$ as control parameters. Our
main technical observation is that the model is exactly solvable
for any value of those parameters in the limit $N\to \infty$
without any need of introducing GVA. Rather, our method is based
on observing that the replicated partition function possesses a
high degree of invariance in the replica space: an arbitrary
simultaneous $O(N)$ rotation of all $n$ replica vectors ${\bf
x}_a$. An efficient method of dealing  with such integrals was
developed recently in the work of one of the authors
\cite{Fyo,FS1} within the framework of the theory of random
matrices. This method allows us to arrive to the effective free
energy functional in the replica space in the most economic, and
we believe elegant way. The subsequent analysis follows the
standard route of using the Parisi Ansatz for the spontaneous
replica symmetry breaking pattern, and the nature of the
low-temperature phase is known to depend very much on the decay
properties of the covariance function of the disorder
potential\cite{MP,Engel}. Actually, we propose a simple
mathematical criterion which allows one to formally discriminate
between the long-ranged and short-ranged disorder, and also
suggests to single out potentials with logarithmically growing
correlations as a separate intermediate class. For all types of
the disorder we provide a detailed derivation of the free energy
functional, the stability analysis, and a thorough description of
the most important features of the resulting phase diagram. The
latter undergoes considerable modifications reflecting a
competition between the confinement provided by the radius $R$ and
the curvature of the parabolic potential $\mu$. For the limiting
case of infinite confinement radius our results faithfully
reproduce all those following from earlier applications of GVA.
This is just another explicit verification of the expected
exactness of GVA in the limit $N\to \infty$.

One more point we find appropriate to mention here is that the
Parisi scenario of replica symmetry breaking for spin glass models
is changing presently its status from a powerful heuristic method of
theoretical physics to an essentially rigorous mathematical
procedure, well-controlled in the case of models of infinite range.
This important change is mainly due to recent seminal results by
Talagrand \cite{Talagrand1,Talagrand},  based on earlier works
 by Guerra \cite{Guerra}, see also interesting works by Aizenman, Sims and
Starr \cite{ASS}. In particular, Talagrand was able to demonstrate
that the equilibrium free energy emerging naturally in the Parisi
scheme of replica symmetry breaking is indeed the correct
thermodynamic limit of the free energy of the paradigmatic
 Sherrington-Kirkpatrick model and, more recently, for the so-called
spherical model of spin glasses, originally studied by one of the
present authors in \cite{sph1}. It is natural to expect that
similar justification should be possible also for other types of
models with glassy thermodynamics, the model of a particle in
random potential being the most natural candidate. However, the
solution of the problem in the framework of GVA does not seem to
be a promising starting point for such a verification. On the
other hand, the expression for the free energy emerging in our
approach is actually very close in its form to those emerging in
the spherical model of spin glasses (see discussion in the end of
the next section). The revealed similarities of our problem to the
spherical spin glass model give strong evidence in favour of
applicability of Talagrand's method for the model under
consideration. We leave a detailed investigation of this issue for
a future work.

 Finally, we hope that our presentation has also a certain
pedagogical value by providing extended description of a few
technical details known to experts, but which are not always easy
to find in the existing literature.

\section{Definition of the model and its formal treatment by replica method.}

As was discussed in the introduction, we consider a classical
particle confined to an impenetrable spherical box of some finite
size $L$. To ensure the non-trivial behaviour in the limit $N\to
\infty$, one has to scale the radius of the sphere with $N$, and
we denote the corresponding domain as $\{D_N:\,{\bf x}^2\le
N\,R^2\}$ \footnote{For some applications, as e.g. \cite{BrayDean}
and \cite{FSW}, it is useful to keep in mind that the volume
$V_N(R)$ of such a sphere in the large $N$ limit behaves
asymptotically as $V_N\approx L_e^N\,\,\mbox{where} \,\,
L_e=\sqrt{2\pi e}\,\, R$.}.

As usual, the main object of interest for us is to calculate the
ensemble average of the free energy
\begin{equation}\label{freendef}
F=-T\,\ln{Z},\quad Z=\int_{D_N} \exp{-\beta{\cal H}(\{\bf x\})}\,
d {\bf x}\,,
\end{equation}
where $\beta=1/T$ stands for the inverse temperature. The average
of the logarithm over the disorder (denoted in the present paper
by angular brackets) is performed with the help of the standard
replica trick, i.e. the formal identity
\begin{equation}\label{replica}
\left\langle\ln{Z}\right\rangle=\lim_{n\to
0}\frac{1}{n}\ln{\left\langle Z^n\right\rangle},\quad
Z^n=\int_{D_N} e^{-\beta\sum_{a=1}^n{\cal H}\{\bf
x_{a}\}}\prod_{a=1}^n d {\bf x}_{a}\,.
\end{equation}

The standard choice of the energy function for this problem is
\begin{equation}\label{fundef}
{\cal H}\{\bf x\}=\frac{\mu}{2}\sum_{k=1}^N x_k^2+V(x_1,...,x_N)
\end{equation}
with $\mu>0$. A random Gaussian-distributed potential $V({\bf x})$
is characterized by zero mean and the variance specified by the
pair correlation function which we choose in the form ensuring
stationarity:
\begin{equation}\label{2}
\left\langle V\left({\bf x}_1\right) \, V\left({\bf
x}_2\right)\right\rangle=N\,f\left(\frac{1}{2N}({\bf x}_1-{\bf
x}_2)^2\right)\,.
\end{equation}

 The previous analysis \cite{MP,Engel} revealed that one should
 essentially distinguish between two rather different situations. The first
describes the case of a short-range correlated disorder
corresponding to functions $f(x)$ vanishing at infinity, with
$f(x)=e^{-x}$ being a typical representative of that class. In the
second case the correlations are long ranged, and at large
distances the potential grows in such a way that:
\begin{equation}\label{2a}
\left\langle \left[V\left({\bf x}_1\right)- V\left({\bf
x}_2\right)\right]^2\right\rangle\propto \left({\bf x}_1-{\bf
x}_2\right)^{2\gamma},
\end{equation}
the exponent $\gamma$  to be chosen in the range $\quad 0< \gamma<
1$. The particular case $\gamma=1/2$ corresponds to a potential
$V({\bf x})$ being the standard Brownian motion. Although formally
$V({\bf x})$ in the latter case  can not satisfy the property
Eq.(\ref{2}) as its variance $\left\langle V^2\left({\bf
x}\right)\right\rangle$ is obviously position-dependent, one can
easily satisfy oneself that such a difference is completely
immaterial for the free energy calculations. As a result, we
always assume for the long-ranged disorder the validity of
Eq.(\ref{2}) with the choice:
\begin{equation}\label{2b}
f(x)=f(0)-g^2x^\gamma,\quad f(0)>0,\quad  0< \gamma< 1.
\end{equation}

In what follows we will be able to formulate a certain criterium
relating the nature of the low-temperature glassy phase of the
model with the shape of the correlation functions $f(x)$, see
Eq.(\ref{con5}) and the discussion around it. According to that
criterium, in a broad class of random potentials with short-range
correlations the glassy phase will be described by the so-called
one-step replica symmetry breaking (1RSB), whereas all long-ranged
potentials Eq(\ref{2b}) are characterized by the full replica
symmetry breaking (FRSB). The criterium also suggests naturally to
single-out as a special case the logarithmically growing
correlations, that is
\begin{equation}\label{2c}
f(x)=-g^2\ln{(x+a^2)}, \quad a^2<1.
\end{equation}
We shall see that such a choice leads to the phase diagram which
combines some features typical for the short-ranged behaviour and
others for the long-ranged types of disorder. In this sense the
logarithmically growing correlations should be considered as a
marginal case intermediate between the two broad classes described
above. It is interesting to mention that the choice of a glassy
model with logarithmically growing correlations Eq.(\ref{2c}) is
not purely academic, as such objects actually emerge e.g. in a
context of statistics of the wave function in disordered
two-dimensional systems \cite{2d}.

 Actually, a few initial steps in evaluation of the free energy
  are precisely the same for all
 types of disorder, provided the latter is of the Gaussian nature.
 Performing the averaging over the disorder in
Eq.(\ref{replica}), we in the standard way arrive at the following
expression:
\begin{equation}\label{replica1}
\left\langle
Z^n\right\rangle=e^{\frac{\beta^2}{2}Nnf(0)}\int_{D_N} \exp{-\beta
H_n \{{\bf x}_a\}}\prod_{a=1}^n d {\bf x}_{a}\,,
\end{equation}
where
\begin{equation}\label{repham}
 \quad H_n \{{\bf
x_a}\}=\frac{\mu}{2}\sum_{a=1}^n{\bf x}^2_{a}-N\beta\sum_{a<b}
f\left(\frac{1}{2N}({\bf x}_{a}-{\bf x}_b)^2\right)\,.
\end{equation}
So far all our manupulations were exact. To achieve further
progress one has to suggest an efficient way of working with the
resulting multidimensional integral. In the standard model with
infinite box radius $R=\infty$ Mezard and Parisi\cite{MP}
suggested to deal with apparently non-Gaussian character of the
integrand by replacing the exact replicated Hamiltonian $ H_n
\{{\bf x_a}\}$ with a trial Hamiltonian $ H^{(t)}_n \{{\bf x_a}\}$
chosen to be Gaussian with respect to all variables ${\bf x_a}$,
and then to apply a kind of variational principle to find the best
possible Gaussian approximation.

Here we point out the possibility of a different route, valid for
any value of the parameter $R$ and requiring at this step no
approximation.  It is based on observing that the integrand in
Eq.(\ref{replica1}) in fact possesses a high degree of invariance:
it depends on $N-$component vectors ${\bf x_a}$ only via
$n(n+1)/2$ scalar products $q_{ab}={\bf x_a}{\bf x_b},\,\, a\le
b$, and is therefore invariant with respect to an arbitrary
simultaneous $O(N)$ rotation of all vectors ${\bf x}_a$. Moreover,
our choice of the integration domain respects this invariance.

An efficient method of dealing  with such integrals is based on
the possibility of rewriting the integral
\begin{equation}\label{defint}
J_{N,n}=\int_{|{\bf x}_1|<L}... \int_{|{\bf x}_n|<L} {\cal
I}_x({\bf x}_1,...,{\bf x}_n)\,d{\bf x}_1\ldots d{\bf x}_n
\end{equation}
whose integrand ${\cal I}_x({\bf x}_1,...,{\bf x}_n)$ possesses such
type of invariance in an alternative form as
\begin{equation}\label{trans1}
J_{N,n}={\cal C}_{N,n} \int_{D_N^{(Q)}}{\cal I}_Q(Q)\,
\mbox{det}Q^{(N-n-1)/2}\, dQ\,,
\end{equation}
provided $N\ge n+1$. Here in Eq.(\ref{trans1}) the original
integrand ${\cal I}_x(\{{\bf x}_a\}))$ is expressed as a function
${\cal I}_Q(Q)$ of $n\times n$ real symmetric positive
semidefinite matrix $Q\ge 0$ whose entries are precisely those
$q_{ab}$, introduced above. The integration domain $D_N^{(Q)}$ is
simply $D_N^{(Q)}=\{Q\ge 0,\, q_{aa}\le N R^2,\, a=1,\ldots n\}$,
the volume element is $dQ=\prod_{a\le b} dq_{ab}$ and the
proportionality constant is given explicitly by
\begin{equation}\label{const}
{\cal C}_{N,n}=
\frac{\pi^{\frac{n}{2}\left(N-\frac{n-1}{2}\right)}}
{\prod_{k=0}^{n-1}\Gamma\left(\frac{N-k}{2}\right)}
\end{equation}
Although some use of similar formulae in statistical mechanics can
be traced back to \cite{Duplantier}, the full potential of the
transformation from Eq.(\ref{defint}) to Eq.(\ref{trans1}) seems to
be revealed in the work of one of the authors \cite{Fyo,FS1} where
it was rediscovered in the context of random matrix theory using
Ingham-Siegel matrix integrals. Since then the formula found
applications in physics of disordered systems, see e.g \cite{rigor}
for an elegant derivation and further use.

 Applying this procedure to our case and using the subsequent rescaling $Q\to NQ$
yields the following exact expression for the averaged replicated
partition function:
\begin{equation}\label{replica2}
\left\langle Z^n\right\rangle={\cal C}_{N,n} N^{Nn/2}
e^{\frac{\beta^2}{2}Nnf(0)}\int_{D_Q}
\left(\mbox{det}Q\right)^{-(n+1)/2} e^{-\beta N\Phi_n (Q) }\, dQ
\end{equation}
where
\begin{equation}\label{repham1}
 \Phi_n (Q)=\frac{\mu}{2}\sum_{a=1}^n q_{aa}-
 \frac{1}{2\beta}\ln{(\det{Q})}-\beta\sum_{a<b}
f\left[\frac{1}{2}(q_{aa}+q_{bb})-q_{ab}\right]
\end{equation}
and $N$ is assumed to satisfy the constraint $N>n$. The final
integration domain $D_Q$ is already $N-$independent: $D_Q=\{Q\ge
0,\, q_{aa}\le \,R^2,\, a=1,\ldots n\}$.

The form of the integrand in Eq.(\ref{replica2}) is precisely one
required for the possibility of evaluating the replicated
partition function in the limit $N\to \infty$ by the Laplace
("saddle-point") method\footnote{One should note that as long as
one is interested only in finding the leading exponential factors
in $N\to \infty$ limit, one can in principle follow a different
route. Namely, impose $n(n+1)/2$ constraints
$Q_{ab}=\frac{1}{N}{\bf x_a}{\bf x_b}$ through the integral
Fourier representations involving $n(n+1)/2$ auxiliary fields
$\lambda_{ab}$, and take the saddle-point both in $\lambda-$ and
$Q-$ variables. The same asymptotic result Eq.(\ref{repfreeen})
then follows after simple manipulations. We however believe that
the use of the elegant mathematical procedure based on
Eq.(\ref{trans1}) is more conceptually clear, and appealing
aesthetically. We hope that it may also provide a useful basis for
calculating $1/N$ corrections, and perhaps for a rigorous
mathematical treatment of the problem.}. Taking into account the
expressions Eqs.(\ref{freendef}), (\ref{replica}), and
(\ref{const}) the free energy of our model is then given by
\begin{equation}\label{repfreeen}
F_{\infty}=\lim_{N\to \infty}\frac{1}{N}\langle
F\rangle=-\frac{T}{2}\ln(2\pi e)-\frac{1}{2T}f(0)+\lim_{n\to
0}\frac{1}{n}\Phi_n (Q)
\end{equation}
where the entries of the matrix $Q$ are chosen to satisfy the
conditions: $\frac{\partial \Phi_n(Q)}{\partial q_{ab}}=0$ for
$a\le b$. This yields, in general, the system of $n(n+1)/2$
equations:
\begin{equation}\label{sp1}
\mu-\frac{1}{\beta}\left[Q^{-1}\right]_{aa}-\beta\sum_{b(\ne a)}^n
f'\left[\frac{1}{2}(q_{aa}+q_{bb})-q_{ab}\right]=0,\quad
a=1,2,\ldots,n
\end{equation}
and
\begin{equation}\label{sp2}
 -\frac{1}{\beta}\left[Q^{-1}\right]_{ab}+\beta
f'\left[\frac{1}{2}(q_{aa}+q_{bb})-q_{ab}\right]=0,\quad a\ne b
\end{equation}
where $f'(x)$ stands for the derivative $df/dx$.

 One should also ensure that the solutions to these equations
respects the constraint $q_{aa}\le R^2$ for all $a=1,\ldots,n$
imposed by presence of the boundaries of the integration domain
$D_Q$. We will shortly see that in the replica limit $n\to 0$ that
condition will be violated for some regions of parameters $T,\mu$.
If this happens, Eqs.(\ref{sp1}) should be simply replaced by
equalities $q_{aa}=R^2$, and the solution to the remaining set
Eq.(\ref{sp2}) should be sought with those constraints imposed.
One then notices that for $R=1$ the resulting expression for the
free energy formally coincides , up to a trivial constant term,
with that obtained for the so-called "spherical" mean field model
of spin glasses. Various features of the latter model attracted a
lot of research interest in recent years, see e.g.
\cite{sph1,sph2,sph3,Talagrand}.

The correspondence between the free energies of the two models
which is so apparent in our approach deserves a short comment. In
one of its recent incarnations the Hamiltonian ${\cal H}(\sigma)$
of the spherical model was defined as \cite{Talagrand}
\begin{equation}\label{spherical}
{\cal H}({\bf \sigma})=\sum_{p\ge
1}\frac{J_p}{N^{(p-1)/2}}\sum_{i_1,\ldots,i_p}
g_{i_1,\ldots,i_p}\sigma_{i_1}\ldots \sigma_{i_p}\,
\end{equation}
in terms of $N-$component vectors ${\bf \sigma}$ spanning the
sphere of radius $\sqrt{N}$. Here $g_{i_1,\ldots,i_p}$ denote
independent Gaussian variables with mean zero and unit variance.
The original version \cite{sph1} of the same model can be shown to
yield precisely the same free energy in the thermodynamic
limit\cite{Talagrand}. Immediate consequence of the definition
Eq.(\ref{spherical}) is that the correlation function
$\frac{1}{N}\langle {\cal H}({\bf \sigma_1}){\cal H}({\bf
\sigma_2})\rangle_g$ depends on the vectors ${\bf \sigma}_1$ and
${\bf \sigma}_2$ only via the scalar product $({\bf \sigma}_1{\bf
\sigma}_2)$. This is precisely the property sufficient to ensure
global $O(N)$ invariance of the replicated partition function of
the model, and our method of deriving the mean free energy goes
through without any modifications, and may well be the shortest
possible. The spherical constraint $\sum_{i\le N} \sigma_i^2=N$
simply translates into the condition $R=1$.

\section{Analysis of the phase diagram of the model within the
replica symmetric ansatz.}

Our procedure of investigating the equations
Eqs.(\ref{sp1},\ref{sp2}) in the replica limit $n\to 0$ will follow
the standard pattern suggested by developments in spin glass theory.
We first seek for the so-called "replica symmetric" solution, and
then investigate its stability in the $(\mu,T)$ plane. When the
replica symmetric solution is found inadequate, it will be replaced
by the hierarchical ("Parisi", or "ultrametric") ansatz for the
matrix elements $q_{ab}$, with various levels of replica symmetry
breaking. Since the full analysis contains a lot of features to be
explained in detail, the reader may wish to cast a regular look at
the resulting phase diagrams fig1, fig.2 and fig.3 in the process of
reading.

The Replica Symmetric Ansatz amounts to searching for a solution
to Eqs.(\ref{sp1},\ref{sp2}) within subspace of matrices $Q$ such
that $q_{aa}=q_d$, for any $a=1,\ldots n$, and $q_{a<b}=q_0$,
subject to the constraints $0<q_0\le q_d\le R^2$. Inverting such a
matrix $Q$ yields again the matrix of the same structure, with the
diagonal entries all given by
\begin{equation}\label{invsym1}
p_d=\frac{q_d+q_0(n-2)}{(q_d-q_0)(q_d+q_0(n-1))}
\end{equation}
and off-diagonal entries given by
\begin{equation}\label{invsym2}
p_0=-\frac{q_0}{(q_d-q_0)(q_d+q_0(n-1))}
\end{equation}
Note, that
\begin{equation}\label{invsym}
p_d-p_0=\frac{1}{q_d-q_0}
\end{equation}
The system Eqs.(\ref{sp1},\ref{sp2}) is reduced in this way to two
equations, which we write directly in the replica limit $n\to 0$
as
\begin{equation}\label{sp1sym}
\mu-T\,p_d+\frac{1}{T} f'\left(q_{d}-q_{0}\right)=0
\end{equation}
\begin{equation}\label{sp2sym}
 -T\, p_0+\frac{1}{T}
f'\left(q_d-q_0\right)=0
\end{equation}
This system of equations is easy to solve employing the relation
(\ref{invsym}), and to obtain
\begin{equation}\label{qsym}
q_d=\frac{T}{\mu}-\frac{1}{\mu^2}f'\left(\frac{T}{\mu}\right),\quad
q_0=-\frac{1}{\mu^2}f'\left(\frac{T}{\mu}\right)
\end{equation}
In order this solution to be sensible one first of all has to
require $f'(x)<0$, which we always assume to hold in our model. In
addition, the solution Eq.(\ref{qsym}) can hold only as long as
$q_d\le R^2$, which in view of the above expressions amounts to
the condition
\begin{equation}\label{Rdomsym}
R^2-\frac{T}{\mu}+ \frac{1}{\mu^2}f'\left(\frac{T}{\mu}\right)\ge
0 \,\,.
\end{equation}
When the above inequality is violated, we should rather use
$q_d=R^2$, and find $q_0$ from the "spherical model" type equation
\begin{equation}\label{Rdominsym}
\frac{q_0}{(R^2-q_0)^2}+\frac{1}{T^2}f'(R^2-q_0)=0
\end{equation}
following immediately from Eqs.(\ref{sp2sym},\ref{invsym2}).

Let us briefly discuss general properties of the boundary line
$T_b(\mu)$ separating the "$\mu-$dominated" regime from the
"R-dominated" one in $(\mu,T)$ plane, for a fixed value of the
confining radius $R$. For a given value of $\mu$ the value of
$T_b$ is obtained by solving the equation
\begin{equation}\label{Rboundsym}
R^2-\frac{T_b}{\mu}=-
\frac{1}{\mu^2}f'\left(\frac{T_b}{\mu}\right) \,\,.
\end{equation}
When analysing this equation we shall assume in addition to the
condition $f'(x)<0$  two more conditions: the uniform concavity
condition $f''(x)>0$ as well as the condition $f'''(x)<0$.
Sensibility of that choice will be justified by analysis of
patterns of spontaneous replica symmetry breaking in subsequent
sections, see discussions around Eq.(\ref{con5}). We also assume
that $f'(x)\to 0$ for large $x$, as is indeed the case for all
types of disorder in the original model.

For the short-range disorder the values $f'(0)<\infty$ and
$f''(0)<\infty$. It is convenient to define for subsequent use two
quantities
\begin{equation}\label{Rcr}
\mu_{c}=\frac{1}{R}\sqrt{-f'(0)},\quad
R_{cr}=\sqrt{-\frac{f'(0)}{f''(0)}}
\end{equation}
Then a simple graphical analysis of Eq.(\ref{Rboundsym}) shows
that for $\mu>\mu_c$ that equation has a single solution
$T_b(\mu)$ tending asymptotically to $T_b=\mu\,R^2$ for $\mu\gg
\mu_c$. In contrast, for $\mu<\mu_{c}$ the number of solutions
essentially depends on the value of the confining radius $R$.
Defining $R_{cr}$ as in Eq.(\ref{Rcr}), we find that for
$R<R_{cr}$ the condition $\mu<\mu_c$ implies that no such solution
$T_b$ exists at all. This fact corresponds to the picture of
monotonically increasing curve $T_b(\mu)$ starting from the point
$(\mu_c,0)$ in the $(\mu,T)$ plane. In the opposite case
$R>R_{cr}$ there are two solutions $T_{b1}<T_{b2}$ in the whole
interval $\mu_0<\mu\le \mu_c$, with $T_{b2}-T_{b1}\to 0$ as
$\mu\to\mu_0$. The value of $\mu_0$ and the corresponding
temperature value $T_0=T_{b1}=T_{b_2}$ can be found as
\begin{equation}\label{mu0}
\mu_0=\sqrt{f''(\tau_0)}, \quad T_0=\mu_0\tau_0 ,
\end{equation}
 with $\tau_0$ being the solution of the equation
\begin{equation}\label{tau0}
R^2=h(\tau_0),\quad h(\tau)= \tau-\frac{f'(\tau)}{f''(\tau)}.
\end{equation}
Note, that $dh/d\tau=f'(\tau)f'''(\tau)/[f''(\tau_0))]^2>0$
according to our assumptions. Thus, the right-hand side of
Eq.(\ref{tau0}) is a monotonously increasing function, and thus
the equation has a (unique) solution $\tau_0\ge 0$ as long as
$R^2\ge h(0)=R_{cr}^2$. In contrast, for $R<R_{cr}$ the equation
Eq.(\ref{tau0}) has no solutions.

Finally, for $\mu<\mu_0$ the equation (\ref{Rboundsym}) has no
more solutions.

Relegating similar analysis of the long-ranged as well as the
logarithmic correlations to the end of the section, we now discuss
the last important ingredient of the procedure: the issue of the
stability of the emerging solution against fluctuations in the
replica space. Indeed, the very essence of the saddle-point method
calls for a check of the replica symmetric solution being locally
stable, in the sense of corresponding to the true extremum (in the
replica limit $n\to 0$, to a maximum) of the functional $\Phi(Q)$.
This should be judged by analysing the eigenvalues of the matrix
$G_{ab,cd}=\frac{\partial^2\Phi}{\partial q_{ab}\partial q_{cd}}$
at this solution. Such analysis along the lines of the classical
De-Almeida and Thouless (AT) paper\cite{AT} is presented in the
Appendix B. The main outcome is that the $\mu$-dominated replica
symmetric solution Eq.(\ref{qsym}) is locally stable as long as
\begin{equation}\label{ATmu}
\mu^2-f''\left(\frac{T}{\mu}\right)\ge 0,
\end{equation}
whereas the "spherical model"-type solution satisfying
Eq.(\ref{Rdominsym}) is stable provided
\begin{equation}\label{ATspherical}
\frac{1}{(R^2-q_0)^2}-\frac{1}{T^2}f''(R^2-q_0)\ge 0.
\end{equation}
Combining the latter with Eq.(\ref{Rdominsym}) for $q_0$ one
easily finds that the domain of the replica-symmetric stability in
the R-dominated regime is given by
\begin{equation}\label{ATR}
T\ge \tau_0\sqrt{f''(\tau_0)}
\end{equation}
where $\tau_0$ is precisely the solution of Eq.(\ref{tau0}).

We will see shortly that both conditions Eq.(\ref{ATmu}) and
Eq(\ref{ATR}) could be violated for low enough values of $\mu$ and
$T$ below the so-called de-Almeida-Thouless line $T_{AT}(\mu)$.
Actually, in the $R-$dominated region of the phase diagram such a
line is parallel to the $\mu-$axis, as the solution $\tau_0$ of
Eq.(\ref{tau0}) determining the right-hand side of (\ref{ATR})
depends only on $R$ and hence is $\mu-$ independent. In particular,
for the limiting case $\mu=0$ the expression for the whole
de-Almeida-Thouless line in $(R,T)$ plane is just
$T_{AT}(R)=\tau_0\sqrt{f''(\tau_0)}$  (see e.g fig.4(c) for a
particular choice of the short-ranged potential).

Let us see how these features are incorporated into the analysis
of the boundary $T_b(\mu)$ between the $\mu$-dominated and the
$R-$dominated regions performed by us above. In the case of a
short-range disorder with $f''(0)<\infty$ studied above we can
easily see that the de-Almeida Thouless line $T_{AT}(\mu)$ given
by the equality sign in Eq.(\ref{ATmu}) must end up for zero
temperature at the point $\mu_{AT}=\sqrt{f''(0)}$. As a simple
consequence, for small confining radius $R<R_{cr}$, with  $R_{cr}$
again given by Eq.(\ref{Rcr}), the whole instability region falls
outside the domain of validity of $\mu$-dominated solution. To
find whether the replica-symmetric solution is stable one
therefore has to use instead the equation given by the equality
sign in Eq.(\ref{ATspherical}).  As the latter equation actually
does not have at all roots for $R<R_{cr}$, we conclude that for
the present model the replica symmetric solution is always stable
for such values of the confining radius $R$. In short, the small
box size $R<R_{cr}$ precludes possibility of the replica symmetry
breaking, hence the glassy behaviour of the models with
short-ranged correlations, see fig.3a\footnote{Note that in recent
works \cite{BrayDean,FSW} precisely the same lengthscale $R_{cr}$
appeared in the analysis of the "geometric complexity" associated
with the zero-temperature limit of the present model. Namely, the
samples with $R<R_{cr}$ were found to be unable to support the
existence of exponentially many saddle points in their energy
landscape. Such an existence seems to be a necessary feature of a
phase with glassy behaviour\cite{FSW}.}

The situation changes crucially for $R>R_{cr}$. First of all, now
the lines $T_b(\mu)$ and the AT line $T_{AT}(\mu)$ can indeed
intersect each other in $(T,\mu)$ plane. Curiously enough, the
point of intersection turns out to coincide with the point
$(\mu_0,T_0)$, see Eq.(\ref{mu0}), which emerged in the previous
analysis as the "leftmost" point on the curve $T_b(\mu)$, such
that for $\mu<\mu_0$ the equation (\ref{Rboundsym}) has no
solutions. For $\mu>\mu_0$ the branch $T_{b1}$ drops therefore
below the AT line and hence has no meaning. At the same time for
another branch $T_{b2}>T_{AT}$ and hence that solution survives.
We shall see in subsequent sections that our analysis of the phase
with broken replica symmetry will provide us with a modified
expression for the boundary line $T_{b1}(\mu)$ extending from the
point $(\mu_0,T_0)$ down to zero temperature.

For $\mu<\mu_0$ we are in the domain of validity of the
R-dominated solution, and the corresponding AT temperature should
be given by $\mu-$ independent value $T_{AT}$ from Eq.(\ref{ATR}).
The latter expression now makes full sense as the corresponding
solution $\tau_0$ indeed exists. Finally, in view of
Eq.(\ref{mu0}) it is evident that the $\mu-$ independent value of
$T_{AT}$ is simply $T_0$ everywhere in R-dominated phase, so that
the R-dominated and $\mu-$ dominated AT lines indeed meet each
other at the same point $(\mu_0,T_0)$ of the phase diagram, as was
natural to expect.

So far our analysis assumed the case of a short-range disorder.
For the case of long-ranged disorder, Eq.(\ref{2b}), the overall
structure of the line $T_b(\mu)$ is in many respects similar, but
has some peculiarities. In particular, there will be no analogue
of the critical value of the confinement radius $R=R_{cr}$ in this
case. Below we proceed in presenting a brief analysis of the
situation for any value of the exponent $\gamma\in (0,1)$. Using
$f'(x)=-\gamma g^2 x^{\gamma-1}$, the equation Eq.(\ref{Rdomsym})
for the boundary line $T_b(\mu)$ between the $\mu-$dominated and
spherical model like regions in the $(\mu,T)$ diagram can be
written as
\begin{equation}\label{Rlinelong}
F(u)=u^{\frac{2-\gamma}{1-\gamma}}-R^2u+\gamma
\frac{g^2}{\mu^2}=0,\quad \mbox{with}\,\,
u=\left(\frac{T_b}{\mu}\right)^{1-\gamma}
\end{equation}
We see that the function $F(u)$ has its single minimum for $u>0$
at $u=u_{min}=
\left(\frac{1-\gamma}{2-\gamma}\,R^2\right)^{1-\gamma}$, and
$F(0)>0,\,F(u\to\infty)>0$. Hence, the equation $F(u)=0$ has two
positive solutions $u_{1,2}$ as long as $F(u_{min})<0$, and no
solutions if $F(u_{min})>0$. Then a simple calculation shows that
two branches $T_{b1}<T_{b2}$ exist as long as
\begin{equation}\label{mumin}
\mu>\mu_0=\frac{g\gamma^{1/2}(2-\gamma)^{1-\gamma/2}}
{R^{2-\gamma}(1-\gamma)^{\frac{1-\gamma}{2}}}\,.
\end{equation}
At $\mu=\mu_0$ those two branches merge: $T_{b1}=T_{b_2}=T_0$
which implies $u_2=u_1=u_{min}$. The corresponding characteristic
temperature can be found as
\begin{equation}\label{Tmin}
T_0=\mu_0u_{min}^{\frac{1}{1-\gamma}}=\frac{g\gamma^{1/2}(2-\gamma)^{-\gamma/2}R^\gamma}
{(1-\gamma)^{-\frac{1+\gamma}{2}}}\,.
\end{equation}
Finally, we should take the presence of the de Almeida-Thouless
conditions Eqs.(\ref{ATmu},\ref{ATR}) into consideration. Recall
that  for the case of a long-ranged disorder $f''(0)=\infty$ in
contrast to the finite value typical for the short-range case. As a
consequence, the AT line now extends in the $(\mu,T)$ plane to
arbitrary large values of $\mu$ and is explicitly given by the
equation
\begin{equation}\label{ATlong}
T_{AT}=\frac{\mu^{\frac{\gamma}{\gamma-2}}}{\left[g^2\gamma(1-\gamma)\right]^{\frac{1}{\gamma-2}}}\,.
\end{equation}
 It is now easy to verify that the boundary line
$T_b(\mu)$ which is given for $\mu\ge \mu_0$ by the two branches
$T_{b1}$ and $T_{b2}$ intersects AT line precisely at the point
$(\mu_0,T_0)$, with $u_{AT}=(T_{AT}/\mu)^{1-\gamma}$. Moreover,
one can check that $F(u_{AT})< 0$ for $\mu>\mu_0$. This simply
implies that $T_{b1}<T_{AT}<T_{b2}$ for any $\mu>\mu_0$. Hence,
the lower branch $T_{b1}$ should be discarded as falling outside
the domain of validity of the replica symmetric solution. The
analysis of the phase with broken replica symmetry presented in
the subsequent sections of this paper reveals that the line
$T_{b1}(\mu)$ shall be rather replaced by a vertical line
$\mu=\mu_0$ extending from the point $(\mu_0,T_0)$ down to zero
temperature. Finally, for all values $\mu<\mu_0$ the de-Almeida -
Thouless line $T_{AT}(\mu)$ is given by the $\mu-$independent
temperature $T_{AT}=T_0(R)$. In particular, this expression just
provides the de-Almeida-Thouless line in $(R,T)$ plane for the
limiting case $\mu=0$.

At last, we provide the analysis of the replica symmetric solution
for the logarithmic correlations, Eq.(\ref{2c}). In this case both
AT line and the boundaries $T_{b1},T_{b_2}$ can be easily found
explicitly in the whole $\mu-$dominated regime. They are given by
\begin{equation}\label{ATlog}
T_{AT}(\mu)=g-\mu \,a^2,\quad
T_{b}(\mu)=\frac{\mu}{2}\left[R^2-a^2\pm\sqrt{(R^2+a^2)^2-4g^2/\mu^2}\right]\,,
\end{equation}
where the first formula holds for $\mu\le \mu_{AT}=g/a^2$, and in
the second formula the upper sign corresponds to $T_{b2}$, and the
lower to $T_{b1}$. The two branches of $T_b(\mu)$ meet at
$\mu=\mu_0=2g/(R^2+a^2)$ so that there is no $T_b$ solution for
$\mu<\mu_0$. The de-Almeida-Thouless line meets the boundary
$T_b(\mu)$ at precisely $\mu_0$. In order to make such intersection
happen one has to ensure that $\mu_0<\mu_{AT}$, which is possible
only if the confinement radius $R$ exceeds the critical value
$R_{cr}=a$. As long as $R>R_{cr}$ for $\mu>\mu_0$ one can see that
$T_{b1}(\mu)<T_{AT}(\mu)<T_{b2}(\mu)$, so that only the upper branch
$T_{b2}$ makes actually sense. A subsequent analysis of the phase
with broken replica symmetry will again reveal that
 the line $T_{b1}(\mu)$ should be replaced by the vertical line $\mu=\mu_0$
 everywhere in the glassy phase. Finally, for
$\mu<\mu_0$ the de-Almeida-Thouless temperature is given by the
$\mu-$ independent value $T_{AT}=T_0=g\frac{R^2-a^2}{R^2+a^2}$.
Again, for the limiting case $\mu=0$ this expression provides the
de-Almeida-Thouless line in the whole $(R,T)$ plane. In particular,
for $R\to R_{cr}=a$ we have $T_{0}\to 0$, showing that there is no
place for broken replica symmetry in the logarithmic case as long as
$R<R_{cr}$.

\begin{figure}[h!]
\centering
\includegraphics*[width=7.3cm]{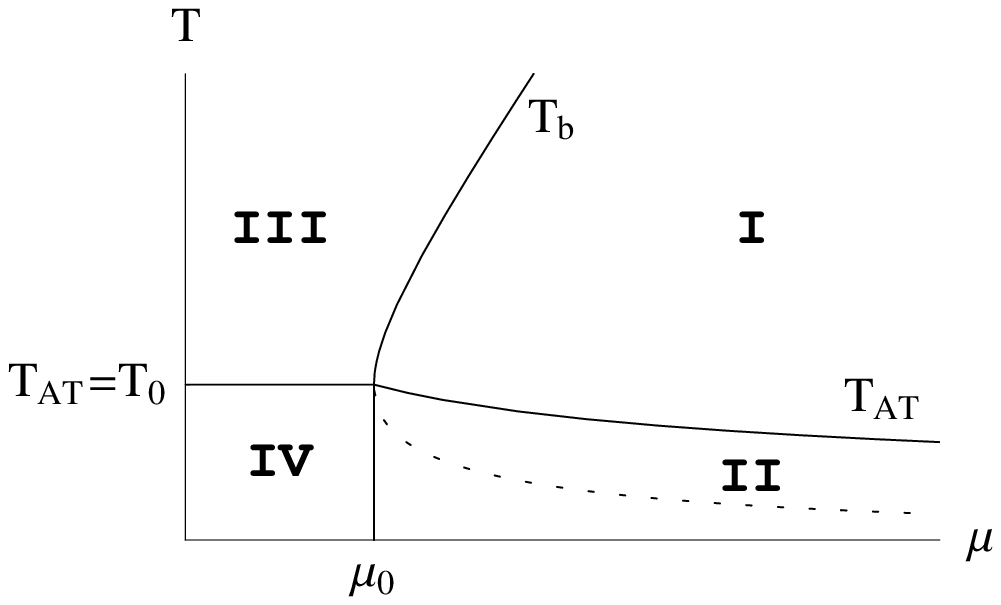}
\caption*{(a)} \hspace*{1.7cm}
\includegraphics*[width=6.9cm]{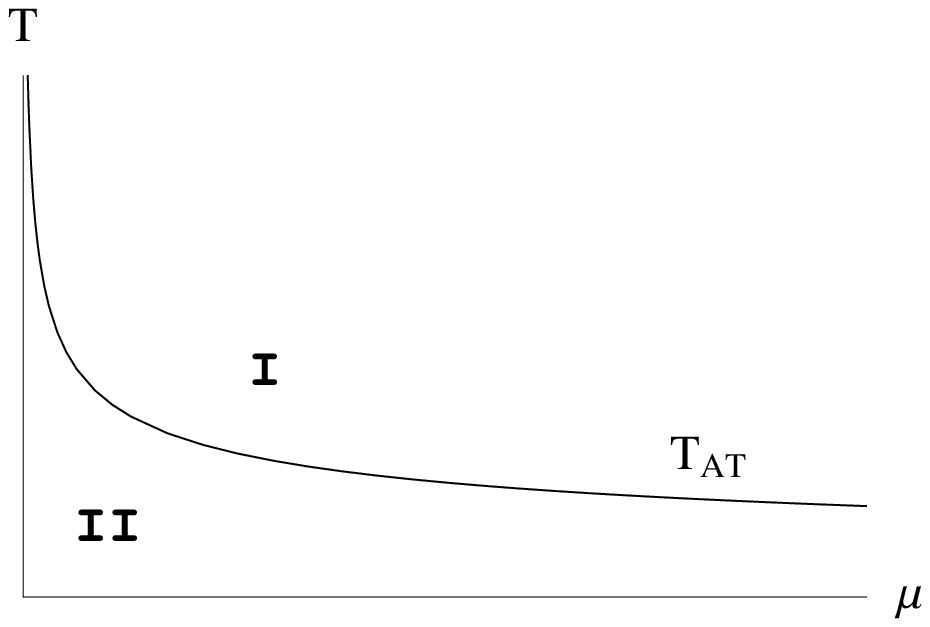}
\caption*{(b)} \hspace*{1.7cm}
\includegraphics*[width=6.9cm]{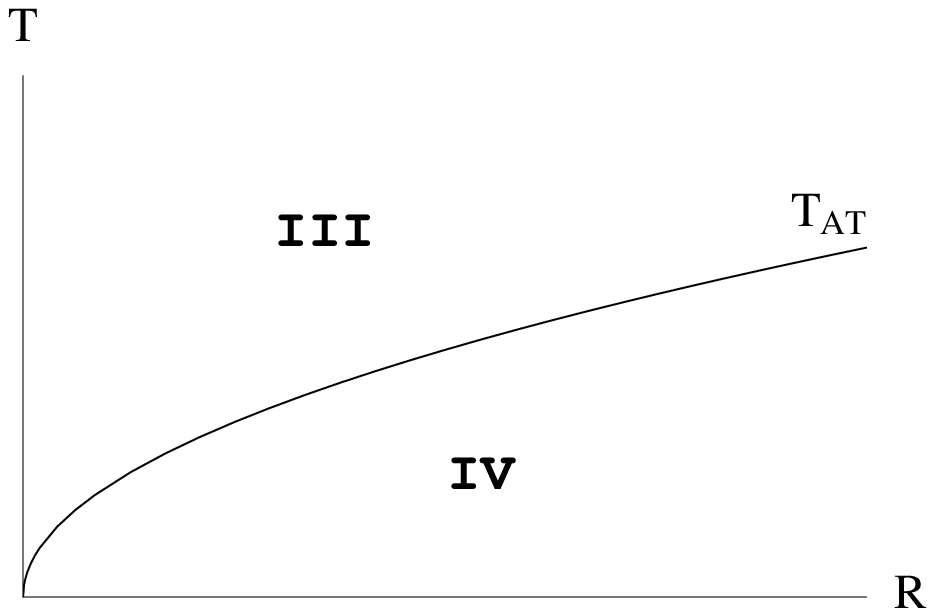}
\caption*{(c)} \caption*{Figure 1: The phase diagrams for the
long-ranged potential with $\gamma=1/2$ and $g=2$. The case (a)
corresponds to $R=\sqrt{3}$, the case (b) to $R=\infty$, and the
case (c) to the choice $\mu=0$. Dotted line in the case (a)
represents the wrong branch of the boundary $T_b$ between the $R-$
dominated and $\mu-$ dominated phases with broken replica symmetry,
and is replaced by the vertical full line. The notation for phases
are as follows: {\bf I} stands for $\mu-$dominated replica-symmetric
(RS) phase; {\bf II} for $\mu-$dominated glassy phase with broken
RS; {\bf III} for $R-$dominated RS phase, and {\bf IV} for
$R-$dominated glassy phase with broken RS;}

\centering
\end{figure}

To summarize our findings, we present in fig.1, fig.2 and fig.3
the resulting phase diagrams in $(\mu,T)$ plane for the particular
choice (i) $f(x)=f(0)-g^2x^{1/2}$ of long-range disorder, (ii) for
the logarithmic correlations, and finally (iii) $f(x)=e^{-x}$
corresponding to the short range disorder. In all these cases the
AT line can be found explicitly. For the $\mu-$dominated regime in
case (i) $T_{AT}=\frac{g^{4/3}}{\mu^{1/3}2^{4/3}}$ and in case
(iii) $T_{AT}=-2\mu\ln{\mu}$. The corresponding coordinates of the
intersection point $(\mu_0,T_0)$ for the two models: (i)
$\mu_0=\frac{3^{3/4}\,g}{2\,R^{3/2}},\,T_0=\frac{g
R^{1/2}}{3^{1/4}\, 2}$ and (iii)
$\mu_0=e^{-(R^2-1)/2},\,T_0=(R^2-1)\, e^{-(R^2-1)/2}$. Note, that
the branch $T_{b1}$ of the solution to (\ref{Rboundsym}) has no
particular meaning as it drops as a whole below the AT line. In
the presented phase diagrams we also included the replacement of
that line by the correct boundaries $T_b(\mu)$ everywhere in the
glassy region $T<T_{AT}$.

\begin{figure}[h!]
\centering \hspace*{1.1cm} \includegraphics[width=6cm]{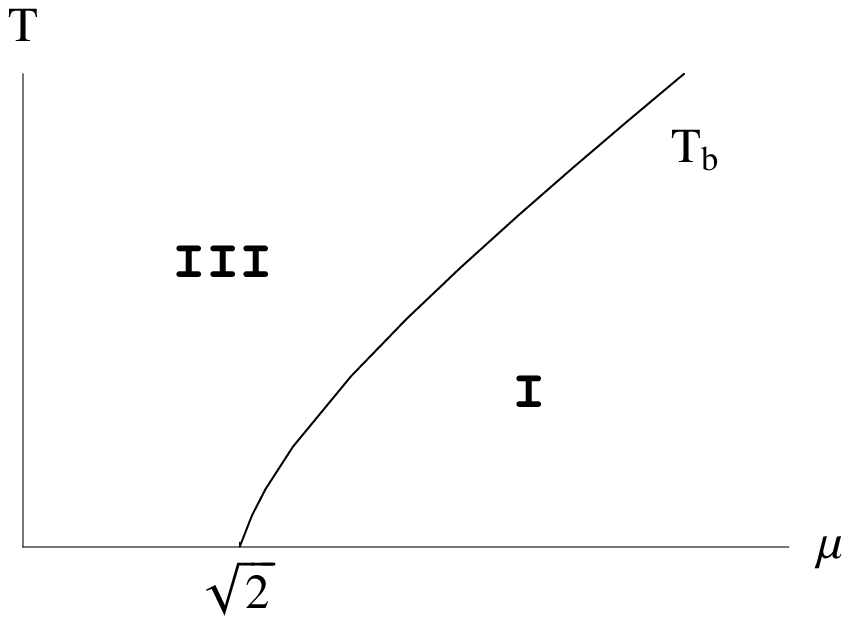}
\caption*{( a)}
\includegraphics[width=6cm]{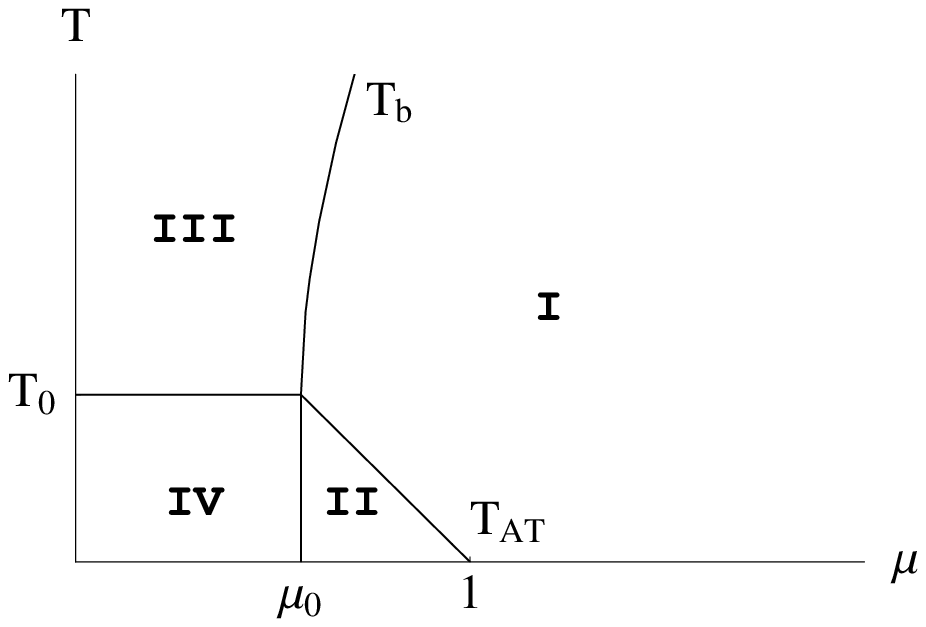}
\caption*{ (b)}
\includegraphics[width=6cm]{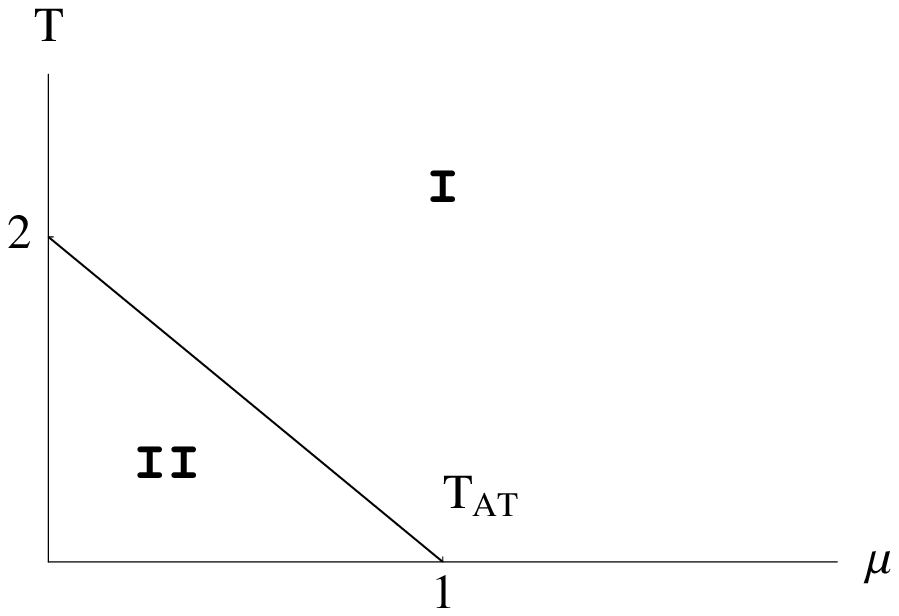}
\caption*{ (c)}
\includegraphics[width=6cm]{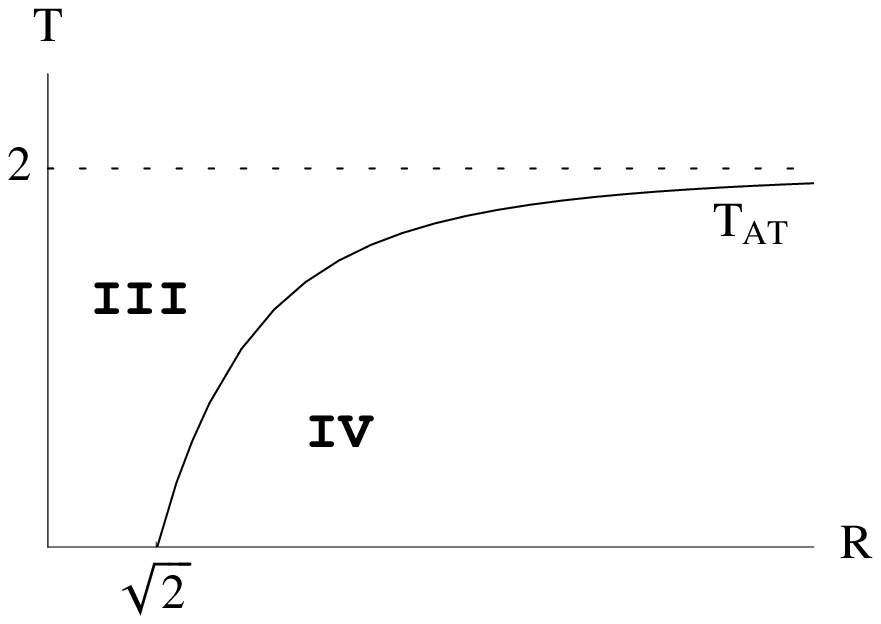}
\caption*{(d)} \caption*{ Figure 2: The phase diagrams for the
potential with logarithmic correlations, with $g=2, a=\sqrt{2}$. In
the case (a) the radius of the box is $R=1<R_{cr}=\sqrt{2}$, in the
case (b) $R=\sqrt{5}>R_{cr}$, in the case (c) $R=\infty$. The case
(d) corresponds to the choice of parameter $\mu=0$. The correct
boundary $T_b$ between the $R-$ dominated and $\mu-$ dominated
glassy phases for the case (b) is represented by full vertical line.
The notation for phases are the same as in fig.1. }


\centering
\end{figure}

We also presented for completeness the typical phase diagrams for
systems with long-range and logarithmic correlations in two limiting
cases: in the $(\mu,T)$ plane for $R=\infty$ and in $(R,T)$ plane
for $\mu=0$. The de-Almeida-Thouless temeperature in the latter case
is equal to $T_{AT}=T_0(R)=\frac{gR^{1/2}}{3^{1/4}\,2}$ for the
long-ranged potential with $\gamma=1/2$. The corresponding diagram
for the case of short-range correlations will be presented in the
end of section Sec.(\ref{l}).

\begin{figure}[h!]
\centering  \hspace*{0.2cm} \includegraphics[width=8cm]{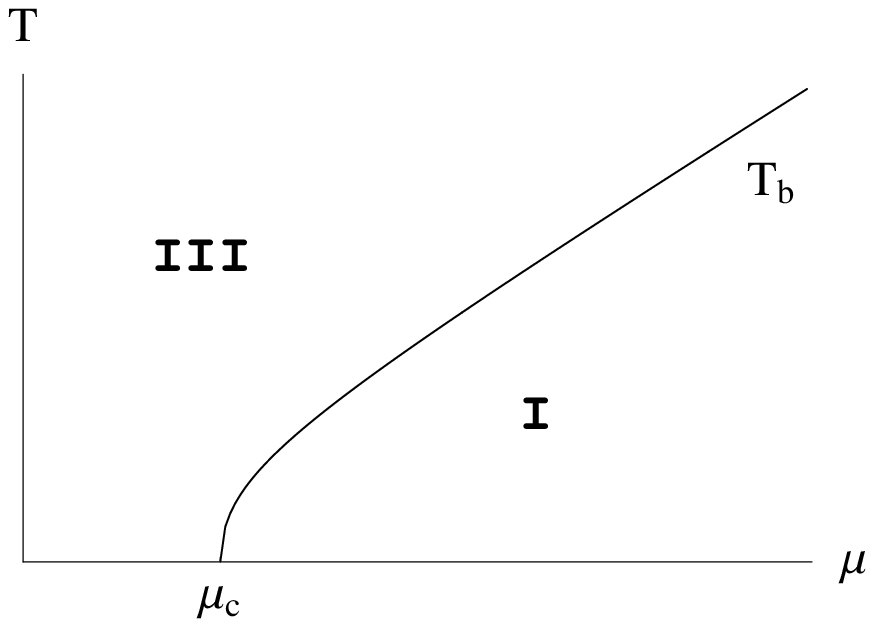}
\caption*{ (a)}
\includegraphics[width=9.3cm]{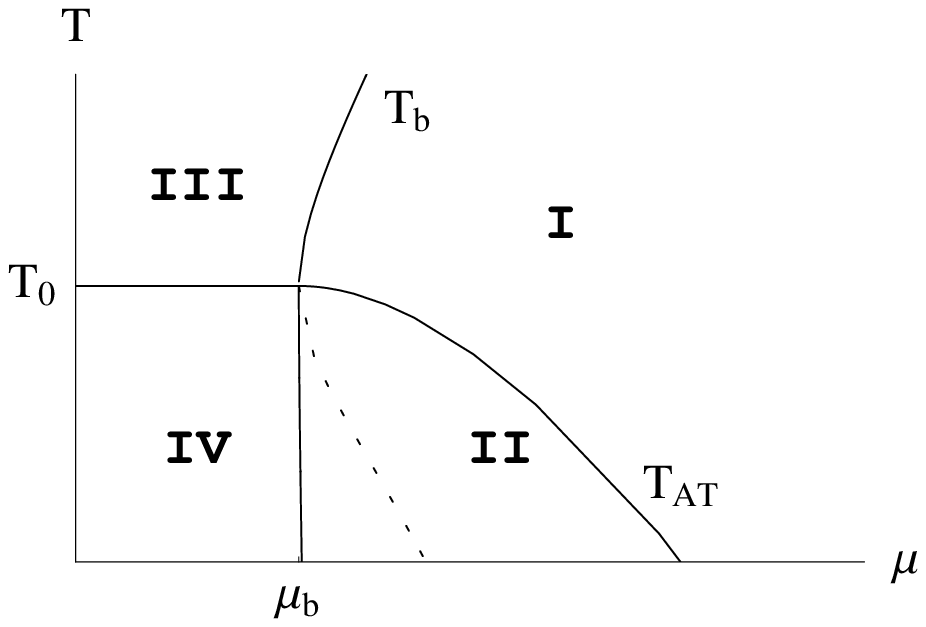}
\caption*{(b)} \caption*{ Figure 3: The phase diagrams for the
potential with short-range correlations, $f(x)=e^{-x}$. In the case
(a) the radius of the box is $R=1=R_{cr}$, in the case (b)
$R=\sqrt{3}=R_{*}$. Dotted line represents the wrong branch of the
boundary $T_b$ between the $R-$ dominated and $\mu-$ dominated
glassy phases and should be replaced by the full line close to
vertical. The notation for phases are the same as in fig. 1. }

\centering
\end{figure}


We finish this section with writing down explicit expressions for
the equilibrium free energy in the replica-symmetric solution.
They are given correspondingly by
\begin{equation}\label{repsymfreemu}
F_{\infty}=-\frac{T}{2}\ln{(2\pi
T/\mu)}-\frac{1}{2T}\left[f(0)-f(T/\mu)\right]
\end{equation}
for the $\mu-$ dominated region; in particular,
$F_{\infty}|_{T=0}=f'(0)/2\mu$. As for the $R-$ dominated RS
region, the free energy is given by
\begin{equation}\label{repsymfreeR}
F_{\infty}=\frac{1}{2}R^2(\mu-\frac{1}{t})-\frac{T}{2}\ln{(2\pi t
T )}-\frac{1}{2T}\left[f(0)-f(tT)\right]\,,
\end{equation}
 with $t$ being the solution of the equation ( see
Eq.(\ref{Rdominsym}))
\begin{equation}\label{Rdominsymtau}
\frac{R^2}{t^2}-\frac{T}{t}+f'(tT)=0\,
\end{equation}
In particular, $t|_{T=0}=R/\sqrt{-f'(0)}$ and
$F_{\infty}|_{T=0}=\frac{1}{2}\mu\,R^2-R\sqrt{-f'(0)}$. Along the
line $T_b(\mu)$ given by Eq.(\ref{Rboundsym}) we obviously have
$t=1/\mu$ and the two expressions for the free energy indeed
coincide. In particular, the two expressions for
$F_{\infty}|_{T=0}$ are indeed equal for $\mu=\mu_c$, see
Eq.(\ref{Rcr}).

\section{Free energy functional within the Parisi scheme of replica symmetry breaking.}

In the region below the AT lines one must discard the unstable
replica symmetric solution in favour of one with the broken
symmetry. In the present section we derive the expression for the
free energy of our model in the phase with broken replica
symmetry. We will follow a particular heuristic scheme of the
replica symmetry breaking proposed originally by Parisi
\cite{Parisi} in the framework of the SK model and employed for
the case of the spherical spin glass in \cite{sph1}.

In our dealing with the free energy functional Eq.(\ref{repham1})
we follow closely the method suggested in the paper by Crisanti
and Sommers \cite{sph1}. To make the present paper self-contained
we choose to describe the procedure in Appendix A and to provide a
few technical details skipped in \cite{sph1}. The analysis is
based on explicit calculation of the eigenvalues of the Parisi
matrix $Q$.

We are actually interested in the replica limit $n\to 0$.
According to the Parisi prescription explained in detail in the
Appendix A, in such a limit the system is characterized by a
non-decreasing function of the variable $q$ denoted as $x(q)$.
Such a function depends non-trivially on its argument in the
interval $q_0\le q \le q_k$, with $q_0\ge 0$ and $q_k\le q_d$.
Outside that interval the function stays constant:
\begin{equation}\label{outside}
x(q<q_0)=0,\quad \mbox {and}\quad  x(q>q_k)=1.
\end{equation}

 In general, the function $x(q)$ also depends on the increasing
sequence
 of $k$ positive parameters $m_i$ satisfying the following
 inequalities
\begin{equation}\label{parisiseq2}
0\le m_1\le m_2\le \ldots\le m_k\le m_{k+1}=1\,.
\end{equation}

As shown in the Appendix A, in the replica limit the following
identity must hold for any differentiable function $g(x)$:
\begin{equation}\label{traces}
\lim_{n\to 0}\frac{1}{n}Tr\left[
g(Q)\right]=g\left(q_d-q_k\right)+
\int_{0}^{q_k}g'\left(\int_{q}^{q_d}x(\tilde{q})\,d\tilde{q}\right)\,dq\,.
\end{equation}

In particular, for the first two terms entering the replica free
energy functional Eq.(\ref{repham1}) application of the rule
Eq.(\ref{traces}) gives
\begin{equation}\label{freeenergy1}
\lim_{n\to 0}\frac{1}{n}\left[\frac{\mu}{2}Tr\, Q
-\frac{1}{2\beta}Tr\ln{(Q)}\right]=\frac{\mu}{2}\,q_d-\frac{1}{2\beta}\ln{(q_d-q_k)}
-\frac{1}{2\beta}\int_{0}^{q_k}\frac{1}{\int_{q}^{q_d}x(\tilde{q})\,d\tilde{q}}\,dq\,.
\end{equation}
The last term in Eq.(\ref{repham1}) is also easily dealt with in
the Parisi scheme (see Appendix A), where it can be written as
\begin{equation}\label{freeenergy2}
\lim_{n\to 0}-\frac{1}{n}\sum_{a\ne b}
f\left[\frac{1}{2}(q_{aa}+q_{bb})-q_{ab}\right]=\lim_{n\to
0}\sum_{l=0}^k(m_{l+1}-m_{l})f(q_d-q_l)=\int_0^{q_d}f(q_d-q)x'(q)\,dq,
\end{equation}
by using explicitly the derivative of the generalized function Eq.
(\ref{xstep}). Using integration by parts and taking into account
the properties Eq.(\ref{outside}) we finally arrive at the
required free energy functional for the phase with broken replica
symmetry
\begin{eqnarray}\label{freebroken}\nonumber
&& F_{\infty}=\frac{\mu}{2}\,q_d-\frac{T}{2}\left[\ln{\left(2\pi e
(q_d-q_k)\right)}+\int_{0}^{q_k}\frac{1}{q_d-q_k+\int_{q}^{q_k}x(\tilde{q})\,d\tilde{q}}\,dq\right]
\\
&&+\frac{1}{2T}\left(f(q_d-q_k)-f(0)+\int_{q_0}^{q_k}f'(q_d-q)x(q)\,dq.\right)
\end{eqnarray}
 This functional should be now extremised with respect to the
non-negative non-decreasing continuous function $x(q)$. In the
$\mu-$dominated regime it also should be extremised with respect
to $q_d$, whereas for the $R-$dominated situation the latter
variable is fixed to be $q_d=R^2$.

\subsection{Solution within the scheme of fully broken replica
symmetry.}

The actual analysis essentially depends on whether we consider the
number of steps $k$ in the Parisi scheme of the replica symmetry
breaking  (see Appendix A) to be finite, or allow it to be
infinite. In the latter situation one conventionally speaks about
a case of full replica symmetry breaking (FRSB), and in the rest
of the present section we restrict our consideration to this
specific case. The number of steps $k$ is assumed to tend to
infinity in such a way, that the sequence of the parameters $m_l$
is replaced with a continuous variable $u\in[0,1]$, such that
$m_{l+1}-m_l=\Delta m_{l}\to du$. The sequence of parameters $q_i$
satisfying Eq.(\ref{parisiseq1}) is simultaneously transformed to
a non-decreasing differentiable function $q(u)$ changing between
its minimal value $q(u=0)=q_0$ and its maximal value $q(u=1)=q_k$.
The function $x(q)$ defined in Eq.(\ref{xstep}) is then assumed to
be transformed to a non-decreasing function continuous in the
interval $q_0\le q \le q_k$, and at least once differentiable
there. Outside that interval it satisfies Eq.(\ref{outside}). As
is easy to see, the identity $x[q(u)]=u$ holds as long as $q_0\le
q(u) \le q_k$.

Variation of the free energy functional Eq.(\ref{freebroken}) with
respect to such a function $x(q)$ as well as with respect to $q_d$
yields
\begin{equation}\label{varfree}
\delta{F_{\infty}}=\frac{1}{2}\left[\int_{q_0}^{q_k}\,\delta x(q)
S(q)dq+\delta q_d\, P\right]=0\,,
\end{equation}
where
\begin{equation}\label{con1}
S(q)=T\int_0^q\frac{d\tilde{q}}{\left[q_d-q_k+\int_{\tilde{q}}^{q_k}x(q)\,d
q\right]^2}+\frac{1}{T}f'(q_d-q)
\end{equation}
and
\begin{equation}\label{con2}
P=\mu-\frac{T}{q_d-q_k}+T\int_0^{q_k}\frac{d\tilde{q}}{[q_d-q_k+\int_{\tilde{q}}^{q_k}x(q)\,d
q]^2}+\frac{1}{T}f'(q_d-q_k)+\frac{1}{T}\int_{q_0}^{q_k}x(q)f''(q_d-q)dq\,.
\end{equation}
Stationarity therefore always amounts to the condition
\begin{equation}\label{stationarity}
S(q)=0,\,\, \forall q\in[q_0,q_k]\,\,,
\end{equation}
and for the $\mu-$ dominated regime we must add also the equation
$P=0$. As $S(q)=0$ implies $\frac{d}{dq}S(q)=0$ we can
differentiate Eq.(\ref{con1}) once, and get after a simple algebra
the equation
\begin{equation}\label{con3}
q_d-q_k+\int_q^{q_k}x(\tilde{q})\,d
\tilde{q}=\frac{T}{\sqrt{f''(q_d-q)}},\quad \forall
q\in[q_0,q_k]\,\,.
\end{equation}
Next differentiation immediately yields the explicit formula for
the function $x(q)$
\begin{equation}\label{con4}
x(q)=-\frac{T}{2}\frac{f'''(q_d-q)}{[f''(q_d-q)]^{3/2}},\quad
\forall q\in[q_0,q_k]\,\,.
\end{equation}
What remains to be determined are the values for parameters
$q_0,q_k,q_d$, the last one only for the $\mu-$ dominated case. To
this end, we substitute the value $q=q_k$ into the relation
Eq.(\ref{con3}) and obtain a simple equation for the difference
$q_d-q_k$:
\begin{equation}\label{condif1}
q_d-q_k=\frac{T}{\sqrt{f''(q_d-q_k)}}\,\,.
\end{equation}
Next, we use the condition $S(q_0)=0$. Exploitation of
Eqs.(\ref{con1},\ref{con3}) yields after a simple algebra one more
relation:
\begin{equation}\label{condif2}
q_0=-\frac{f'(q_d-q_0)}{f''(q_d-q_0)}\,.
\end{equation}
Finally, we can use already obtained relations to simplify
drastically a somewhat cumbersome equation $P=0$, see
Eq.(\ref{con2}), which holds only in the $\mu$-dominated regime.
In particular, the explicit form Eq.(\ref{con4}) for the function
$x(q)$ allows one to write:
\[
\frac{1}{T}\int_{q_0}^{q_k}x(q)f''(q_d-q)dq=-\frac{1}{2}\int_{q_0}^{q_k}\frac{f'''(q_d-q)}{[f''(q_d-q)]^{1/2}}dq=
[f''(q_d-q_k)]^{1/2}-[f''(q_d-q_0)]^{1/2}
\]
Now taking into account the condition $S(q_k)=0$ and the relation
Eq.(\ref{condif1}) allows one to see that the stationarity
condition $P=0$ amounts simply to
\begin{equation}\label{condif3}
\mu^2=f''(q_d-q_0)\,.
\end{equation}
which in turn allows to rewrite Eq.(\ref{condif2}) in the
$\mu-$dominated regime as
\begin{equation}\label{condif2a}
q_0=-\frac{1}{\mu^2}f'(q_d-q_0)\,.
\end{equation}
The system of three equations Eq.(\ref{condif1}),
Eq.(\ref{condif3}) and Eq.(\ref{condif2a}) for the $\mu-$
dominated regime or of the two equations Eq.(\ref{condif1}),
Eq.(\ref{condif2}) for the $R-$ dominated regime allows one to
determine the parameters $q_0,q_k$ (and, if necessary, $q_d$) as
long as the function $f(x)$ is specified. The corresponding
equilibrium free energy can be written, after some algebraic
manipulations exploiting
Eq.(\ref{con3},\ref{con4},\ref{condif1},\ref{condif2}), in the
form valid for both regimes:
\begin{eqnarray}\label{equilibriumfreeglass}
\nonumber && F_{\infty}=\frac{\mu}{2}q_d-\frac{T}{2}\ln{\left[2\pi
e(q_d-q_k)\right]}+\frac{1}{2T}\left[f(q_d-q_k)-f(0)-(q_d-q_k)f'(q_d-q_k)\right]
\\&&
-q_0\sqrt{f''(q_d-q_0)}-\int_{q_0}^{q_k}\sqrt{f''(q_d-q)}\,dq,
\end{eqnarray}
where for the $R-$dominated phase we assume that $q_d=R^2$.

This seemingly completes our solution.

Consistency of the Parisi FRSB scheme requires, however, that the
emerging function $x(q)$ given by Eq.(\ref{con4}) must be a
real-valued non-negative non-decreasing function of its argument.
Non-negativity and reality of this function is indeed ensured by
earlier imposed conditions $f''(x)>0$ and $f'''(x)<0$.
Non-negativity of the derivative of the right-hand side in
Eq.(\ref{con4}) yields as the necessary consistency condition a
new inequality $A(q_d-q)\le 0$, where the function $A(x)$ is
expressed in terms of $f(x)$ as
\begin{equation}\label{con5}
A(x)=\frac{3}{2}\left[f'''(x)\right]^2-f''(x)f''''(x).
\end{equation}
Checking our three main choices for the function $f(x)$ - the
long-ranged, the short ranged and the logarithmic - we observe
that the above inequality is indeed strictly satisfied for any
long-ranged potential Eq.(\ref{2b}) where
\[
A(x)=\frac{1}{2}g^2\gamma^2(\gamma-1)^2(\gamma-2)x^{2(\gamma-3)}<0,\quad
\forall \gamma\in(0,1)\,.
\]
The inequality is only marginally satisfied in the logarithmic
case Eq.(\ref{2c}) where it is easy to see that $A(x)\equiv 0$.
And it is strictly violated in a typical short-range potential, as
e.g. for $f(x)=e^{-x}$ when $A(x)=\frac{1}{2}e^{-2x}>0$. We
therefore conclude that the full replica symmetry breaking can not
occur in short-range models, neither in the $\mu-$ dominated, nor
in $R-$dominated regimes, and should therefore be replaced by a
different scheme. We will discuss the necessary modifications in
the next section. As to the long-ranged potentials, the scheme is
fully legitimate and solutions to the above equations can be
easily found for any value of $\gamma\in (0,1)$. In the
$\mu-$dominated regime they are given by following expressions:
\begin{equation}\label{mulongsolutions1}
q_0=\frac{\mu^{\frac{2}{\gamma-2}}}{(g^2\gamma)^{\frac{1}{\gamma-2}}(1-\gamma)^{\frac{\gamma-1}{\gamma-2}}},\quad
q_0=q_d-\left(\frac{\mu^2}{g^2\gamma(1-\gamma)}\right)^{\frac{1}{\gamma-2}},
\quad
q_k=q_d-\left(\frac{T^2}{g^2\gamma(1-\gamma)}\right)^{\frac{1}{\gamma}}
\end{equation}
and
\begin{equation}\label{mulongsolutions2}
x(q)=\frac{T}{2}\frac{2-\gamma}{[g^2\gamma(1-\gamma)]^{\frac{1}{2}}}\frac{1}{(q_d-q)^{\frac{\gamma}{2}}},
\quad q_0\le q\le q_k\,.
\end{equation}
A few more useful relations follow from those above:
\begin{equation}\label{mulongsolutions3}
q_d=(2-\gamma)q_0,\quad
x(q_0)=\frac{T}{2}\frac{2-\gamma}{\mu^{\frac{\gamma}{\gamma-2}}}\,[g^2\gamma(1-\gamma)]^{\frac{1}{\gamma-2}},
\quad \mbox{and} \quad x(q_k)=1-\frac{\gamma}{2}\,.
\end{equation}
According to the general procedure such a solution makes sense as
long as $q_k\ge q_0$, and it is easy to check that the condition
can be rewritten as $T\le T_{AT}$, where $T_{AT}$ is precisely the
de-Almeida-Thouless temperature for this model given by
Eq.(\ref{ATlong}).

A curious feature of the above solution is that the values of
$q_0,x(q_k)$ and $q_d$ turn out to be temperature-independent
everywhere in the phase with broken replica symmetry. As $q_d$ is
nothing else but the thermodynamic expectation value of the
collective displacement in the original model:
$q_d=\frac{1}{N}\left\langle\overline{{\bf x}^2}\right\rangle$ ,
it is conventional to speak in this case about "freezing" of the
system below the AT line. On the other hand, the same feature
ensures that in this "frozen" or glassy phase the boundary line
$T_b(\mu)$ in the $(\mu,T)$ plane separating the $\mu-$dominated
regime from the $R-$dominated one must be vertical. Indeed, it is
given by the condition $q_d=R^2$ which together with
Eq.(\ref{mulongsolutions1}) simply amounts to $\mu=\mu_0$, with
the value of $\mu_0$ given by Eq.(\ref{mumin}). Everywhere in the
$R-$dominated glassy phase we have
\begin{equation}\label{mulongsolutions4}
x(q)=\frac{T}{2}\frac{2-\gamma}{[g^2\gamma(1-\gamma)]^{\frac{1}{2}}}\frac{1}{(R^2-q)^{\frac{\gamma}{2}}},
\quad q_0\le q\le q_k\,,
\end{equation}
with
\[q_0=\frac{R^2}{2-\gamma},\quad
q_k=R^2-\frac{T^{\frac{2}{\gamma}}}{[g^2\gamma(1-\gamma)]^{\frac{1}{\gamma}}}\,.
\]
These observations complete our investigation of the phase diagram
for the case of long-ranged correlations, full version of which
was presented for the particular case $\gamma=1/2$ in fig.1a.

Now we consider the case of logarithmic correlations,
Eq.(\ref{2c}) assuming $R>a$. As for such a potential $A(x)\equiv
0$ this implies $\frac{d}{dq}x(q)=0$ for $q_0\le q\le q_k$.
Indeed, our general relations
Eqs.(\ref{con4}),(\ref{condif1}),(\ref{condif3}) and
Eq.(\ref{condif2a}) give in this case an especially simple
solution
\begin{equation}\label{logsol}
x(q)=\frac{T}{g},\quad q_0\le q\le q_k;\,\, \mbox{where}\,\,
q_0=\frac{g}{\mu};\,\,q_k=2\frac{g}{\mu}-\frac{a^2}{1-T/g} \,\,\,
\mbox{and}\quad q_d=2\frac{g}{\mu}-a^2.
\end{equation}
for the $\mu-$dominated regime. The consistency condition $q_k\ge
q_0$ is satisfied as long as $T\le T_{AT}$, where $T_{AT}$ is as
expected the de-Almeida-Thouless temperature. For this case
$T_{AT}$ is given by Eq.(\ref{ATlog}). The condition $q_d=R^2$
which defines the boundary $T_b(\mu)$ with the $R-$dominated
glassy regime again amounts to $\mu=\mu_0=2g/(R^2+a^2)$, with the
same value of $\mu_0$ we found earlier for this case. Finally,
everywhere in the $R-$dominated glassy phase occupying the
rectangle $T\le T_0=g\frac{R^2-a^2}{R^2+a^2},\,\, \mu\le \mu_0$ in
the $(\mu,T)$ plane we have:
\begin{equation}\label{glassyRlog}
x(q)=\frac{T}{g},\quad q_0\le q\le q_k;\,\, \mbox{where}\,\,
q_0=\frac{R^2+a^2}{2},\,\,q_k=R^2-a^2\,\frac{T/g}{1-T/g}\,.
\end{equation}
Using these expressions one can find the corresponding equilibrium
free energy $F_{\infty}$. The latter takes especially simple form
 for zero temperature:
\begin{equation}\label{equilfreezero}
F_{\infty}|_{T_=0}=\left\{\begin{array}{l}-\frac{g^2}{2\mu
a^2}\quad \mbox{for}\quad \mu\ge \mu_{AT}= g/a^2\,, \\
-\frac{\mu}{2}a^2-g\ln{\left(\frac{g}{\mu a^2}\right)}\quad
\mbox{for}\quad \mu_0\le\mu\le \mu_{AT}\,,\\
\frac{\mu}{2}R^2-g\left[1+\ln{\frac{(R/a)^2+1}{2}}\right] \quad
\mbox{for}\quad 0\le\mu\le \mu_{0}=2g/(R^2+a^2)\,.
\end{array}
\right.
\end{equation}
Here the upper line corresponds to the $\mu-$dominated RS regime,
second line to the $\mu-$dominated solution with broken RS
symmetry, and third line to $R-$dominated solution with broken RS,
respectively.

The corresponding phase diagram was presented in fig.1b.

\subsection{Short-ranged correlated potentials: solution within the
one-step replica symmetry breaking.}

In this section we will discuss the solution of our problem
pertinent for the potentials with the correlation function $f(x)$
corresponding to the uniformly positive values of the function
$A(x)$ defined in Eq.(\ref{con5}). As we argued above, this
situation is typical for the short-ranged correlated potentials.

We have seen that searching for a solution within the FRSB scheme
when $A(x)>0$ leads to a contradiction. The only remaining
possibility within the Parisi hierarchical ansatz is therefore to
assume that the number $k$ of steps in the replica breaking
hierarchy is finite: $1\le k<\infty$. We shall see, adopting for our
model the line of reasoning suggested first by Crisanti and Sommers
in \cite{sph1}, that the condition $A(x)>0$ forces us to select the
value $k=1$ as the only possible. This situation is conventionally
called the one-step replica symmetry breaking (1RSB). In the
Appendix B we also sketch the stability analysis showing that 1RSB
solution is indeed stable versus small fluctuations in the full
replica space.

In the case of a general finite $k$ the function $x(q)$ is again a
non-negative and non-decreasing, but has only a finite number of
points of growth in the interval $[q_0,q_k]$, parametrised by the
sequences of $q_i$ and $m_i\in[0,1]$. The variation of the free
energy with respect to those parameters is again given formally by
the same formula Eq.(\ref{varfree}) but with understanding that
\begin{equation}\label{varx}
\delta x(q)=\sum_{l=0}^k(\delta m_{l+1}-\delta m_l)\theta(q-q_l)-
\sum_{l=0}^k( m_{l+1}- m_l)\delta(q-q_l)\delta q_l\,.
\end{equation}

The corresponding part of the variation of the free energy is
therefore given by
\[
\delta F=\frac{1}{2}\sum_{l=0}^k(\delta m_{l+1}-\delta
m_l)\int_{q_i}^{q_k}S(q)\,dq- \frac{1}{2}\sum_{l=0}^k( m_{l+1}-
m_l)\,S(q_i)\delta q_l\,,
\]
where the function $S(q)$ is defined in Eq.(\ref{con1}). As
$\delta m_0=\delta m_{k+1}=0$, the above expression can be
represented in the form
\begin{equation}\label{varF}
2\delta F=\sum_{l=1}^k\delta m_{l}\int_{q_{l-1}}^{q_l}S(q)\,dq-
\sum_{l=0}^k( m_{l+1}- m_l)\,S(q_i)\delta q_l\,,
\end{equation}
showing that the stationarity conditions amount to the system of
equations
\begin{equation}\label{statkRSB}
\int_{q_{l-1}}^{q_l}S(q)\,dq=0\,\,\mbox{for}\,\,l=1,2,\ldots,k-1,\,\,\mbox{and}\,\,S(q_l)=0
\,\,\mbox{for}\,\,l=0,1,\ldots,k\,.
\end{equation}
As the function $S(q)$ is obviously continuous in $[q_0,q_k]$, the
first type of condition ensures that it takes both positive and
negative values in each of the intervals $[q_{l-1},q_l]$. According
to the second condition $S(q)$ vanishes at both ends of those
intervals, therefore this function must have at least one maximum
and at least one minimum in each of the above intervals. Denoting
position of those extrema as $q_e$, we must have
$\frac{d}{dq}S(q)|_{q_{e}}=0$ for each point of extremum. Then
differentiating Eq.(\ref{con1}) we get after a simple algebra the
equation whose solutions should give us all possible positions of
extrema $q_e$:
\begin{equation}\label{con33}
q_d-q_k+\int_{q_e}^{q_k}x(\tilde{q})\,d
\tilde{q}=\frac{T}{\sqrt{f''(q_d-q_e)}},\quad \forall
q_e\in[q_0,q_k]\,.
\end{equation}
Now, it is convenient to consider both the right- and left-hand side
of this relation as some functions of the variable $q_e$. Taking the
derivative over $q_e$ from the left-hand side once gives
$-x(q_e)<0$, showing that the left-hand side is a decreasing
function of its argument. Next differentiation yields
$-\frac{d}{dq_e}x(q_e)\le 0$, which shows that the left hand side is
a {\it concave} decreasing function. Now we treat the right-hand
side of Eq.(\ref{con33}) in the same way. First differentiation
shows that the right-hand side is also a decreasing function of
$q_e$ due to the condition $f'''(x)$. At the same time its second
derivative turns out to be equal to
\[
\frac{T}{2}\,A(q_d-q_e)\,\frac{1}{[f''(q_d-q_e)^{5/2}}>0
\]
due to the condition $A(x)>0$. Thus we see that the right-hand
side is a {\it convex} decreasing function of $q_e$. As any convex
decreasing function can have at most two points of intersection
with a concave decreasing one, there is no more than two different
solutions of the Eq.(\ref{con33}) for $q_e$. Hence, there could be
only a single interval $[q_0,q_1]$ in the description of the
function $x(q)$. Using henceforth the notation $m_1\equiv m$, we
see that such a function is given simply by
\begin{equation}\label{x1RSB}
x(q)=\left\{\begin{array}{l} 0, \quad q<q_0\\ m,\,\,\, q\in [q_0,q_1]\\
1,\quad q>q_1 \end{array}\right.\,.
\end{equation}
This is precisely the one-step replica symmetry breaking (1RSB)
hierarchical ansatz, which is thus shown to be the only possibility
within the Parisi scheme for systems satisfying $A(x)>0$.

In the rest of this section we study the phase diagram resulting
from the implementation of such a 1RSB scheme in our model.
Instead of using the general stationarity conditions
Eq.(\ref{statkRSB}), we prefer to start directly with the
variational free energy functional depending on the parameters
$q_0,q_1,q_d$ and $m\in[0,1]$. In fact, we find it more convenient
to use the set of variables
\[m;\,q_0;\, Q=q_1-q_0;\,\,\mbox{and} \,\, y=q_d-q_1\] as independent
variational parameters. Substituting Eq.(\ref{x1RSB}) into
Eq.(\ref{freebroken}) and using the above notations we arrive at
the variational free energy of the form
\begin{eqnarray}\label{free1RSBa}\nonumber
&& F^{1RSB}_{\infty}=\frac{\mu}{2}\,(q_0+y+Q)-\frac{T}{2}\ln{(2\pi
e)}-\frac{T}{2}
\left(1-\frac{1}{m}\right)\ln{y}-\frac{T}{2m}\ln{(y+m\,Q)}
\\ &&
-\frac{T}{2}\frac{q_0}{y+m\,Q}+\frac{1}{2T}\left[(1-m)f(y)+m\,f(y+Q)-f(0)\right]\,.
\end{eqnarray}
One should also remember that in the $\mu-$dominated regime the
found solution should respect the inequality
\begin{equation}\label{Rconstraint}
q_0+y+Q=q_d\le R^2\,.
\end{equation}
In the $R-$dominated regime the above inequality must be replaced
with the equality, and the arising constraint reduces the number
of independent variational parameters to three.

Let us start first with the $\mu-$dominated regime. The parameter
$q_0$ enters in the functional linearly, and can be immediately
excluded in favour of the relation
\begin{equation}\label{muconstraint}
y=y_m(Q)=\frac{T}{\mu}-mQ
\end{equation}
This fact allows one to write down the variational free energy as
the function of only two variables, $m$ and $Q$. It is also
natural to operate with the so-called excess free energy given by
the difference $\Delta F=F^{1RSB}_{\infty}-F^{RS}_{\infty}$
between the free energy value of the RS solution,
Eq.(\ref{repsymfreemu}) and the variational $1RSB$ free energy,
 Eq.(\ref{free1RSBa}). After simple manupulations we obtain
\begin{eqnarray}\nonumber
&& \Delta F=\frac{\mu}{2}(1-m)Q+\frac{T}{2}\,
\frac{1-m}{m}\ln{\left(1-\frac{\mu}{T}\,mQ\right)}
\\ && \label{free1RSBb}
+\frac{1}{2T}\left[(1-m)f\left(\frac{T}{\mu}-mQ\right)+
m\,f\left(\frac{T}{\mu}+(1-m)Q\right)-f\left(\frac{T}{\mu}\right)\right]\,.
\end{eqnarray}
Note that for $m=1$ the difference $\Delta F$ vanishes, since this
choice obviously brings us back to the replica-symmetric solution.
Differentiating Eq.(\ref{free1RSBb}) over $Q$ and assuming
$(1-m)\ne 0$ we obtain the first stationarity condition
\begin{equation}\label{Q1RSBa}
Q= \frac{y_m(Q)}{\mu
T}\left(f'\left[y_0(Q)\right]-f'\left[y_m(Q)\right]\right)\,,
\end{equation}
where $y_m(Q)$ is the combination from the right-hand side of
Eq.(\ref{muconstraint}), and we introduced also the notation
$y_0(Q)=y_m(Q)+Q$. Similarly differentiation of
Eq.(\ref{free1RSBb}) over $m$ yields another equation
\begin{equation}\label{m1RSBa}
-\frac{T}{m^2}\ln{\left[\frac{\mu}{T}y_m(Q)\right]}=
Q\left(\frac{\mu}{m}+\frac{1}{T}f'\left[y_0(Q)\right]\right)-\frac{1}{T}
\left(f\left[y_0(Q)\right]-f\left[y_m(Q)\right]\right).
\end{equation}
 This set of equations determines the system behaviour in the
$\mu-$dominated regime of the phase with one-step broken replica
symmetry. Finally, to find the transition line $T_b(\mu)$ to
$R-$dominated regime one needs to check the condition
Eq.(\ref{Rconstraint}). For this one should be able to express the
value of $q_0$ in terms of $Q$ and $m$.  To this end we notice
that the stationarity condition with respect to $Q$ at the level
of the original free energy expression Eq.(\ref{free1RSBa})
together with Eq.(\ref{muconstraint}) immediately produces the
required relation:
\begin{equation}\label{q01RSBa}
q_0= -\frac{1}{\mu^2}f'\left[y_0(Q)\right]\,.
\end{equation}

Before discussing features of the resulting phase diagram in full
generality it is worth pointing out  the existence of a particular
case when the system of equations Eqs.(\ref{Q1RSBa},\ref{m1RSBa})
and Eq.(\ref{muconstraint}) allows for an explicit algebraic
solution. This is precisely the case of logarithmically correlated
potential, Eq.(\ref{2c}). Indeed, a direct verification shows that
the substitution
\begin{equation}\label{RSB1log}
Q=\frac{g}{\mu}-\frac{a^2}{1-T/g}, \quad m=\frac{T}{g}\,.
\end{equation}
solves all those equations. This observation should look less
surprising if one comes back to the solution Eq.(\ref{logsol})
found earlier as a particular limiting case of the full replica
symmetry breaking ansatz. One then realizes that the corresponding
function $x(q)$ was constant in the interval $q\in [q_0,q_k]$, the
feature being characteristic rather of one-step replica symmetry
breaking. This is just another evidence towards marginal nature of
the potential with logarithmic correlations: the case can be
looked at both as a limiting special case of FRSB solution, and
that of 1RSB solution.

After this digression, we proceed to discussing typical features
of the phase diagram in a generic case of one-step replica
symmetry breaking. As we have already seen, the transition to the
phase with broken replica symmetry may occur in two different
ways: either along the line $Q\to 0$, or along the line $m\to 1$.
Let us first investigate the behaviour of the free energy
difference (\ref{free1RSBb}) for small $Q\ll 1$. Expanding that
expression to up to the first two nonvanishing terms gives
\begin{equation}\label{freeQsmall}
\Delta F=
\frac{m(1-m)}{2T}\left\{\frac{Q^2}{2}\left[-\mu^2+f''(T/\mu)\right]+\frac{Q^3}{3}
\left[-m\frac{\mu^3}{T}+(\frac{1}{2}-m)f'''(T/\mu)\right]\right\}\,.
\end{equation}
Maximization with respect to $Q$ shows that $Q=0$ for $\mu^2\ge
f''(T/\mu)$ which is precisely the de-Almeida-Thouless condition
Eq.(\ref{ATmu}) of stability of the replica-symmetric solution.
Below the AT line the maximum of the excess free energy happens at
a nonzero value of $Q$ given by:
\begin{equation}\label{Qsmall}
Q=\frac{f''(T/\mu)-\mu^2}{-m\frac{\mu^3}{T}+(\frac{1}{2}-m)f'''(T/\mu)}.
\end{equation}
This allows one to write $\Delta F$ in terms of $m$ only as
\begin{equation}\label{freeQsmalla}
\Delta F=
\frac{m(1-m)}{6T}\frac{\left[-\mu^2+f''(T/\mu)\right]^{3}}
{\left[m\frac{\mu^3}{T}-(\frac{1}{2}-m)f'''(T/\mu)\right]^2}\,.
\end{equation}
We see that the excess free energy grows cubically in the glassy
phase in the vicinity of the AT line. Such a behaviour is typical
for continuous glass transitions to a phase with broken replica
symmetry.

Requiring maximum of this excess free energy with respect to $m$
one finds the equilibrium value of this parameter in the vicinity
of the AT line:
\begin{equation}\label{mQsmall}
m=-\frac{1}{2}\frac{T}{\mu^3}f'''\left(\frac{T}{\mu}\right)\,.
\end{equation}

Such value of $m$ may tend to the maximal possible value $m=1$
when the temperature $T$ and the parameter $\mu$ approach along
the AT line a point $(\mu_m,T_m)$ where they satisfy the system of
two equations:
\begin{equation}\label{Tc}
\frac{\mu_m^3}{T_m}+\frac{1}{2}f'''\left(\frac{T_m}{\mu_m}\right)=0,
\quad -\mu_m^2+f''\left(\frac{T_m}{\mu_m}\right)=0\,.
\end{equation}
It is easy to check that these conditions are precisely those
ensuring that $(\mu_m,T_m)$ is the point of a maximum of the AT
line, i.e. $\frac{dT_{AT}}{d\mu}=0,\,\frac{d^2T_{AT}}{d\mu^2}<0$.
Close to the point of maximum the AT line is described by
\begin{equation}\label{A7}
\tau_{AT}=\frac{3\delta^2}{2}\left[1-\frac{T_m^2}{6\mu_m^4}f''''\left(\frac{T_m}{\mu_m}\right)\right],\quad
\delta\ge 0,
\end{equation}
where $\tau_{AT}=(T_m-T_{AT})/T_m\ll
1,\,\,\delta=(\mu-\mu_m)/\mu_m\ll 1$.

For the point of maximum to happen within the $\mu-$dominated
regime the corresponding value $\mu_m$ should obviously satisfy
$\mu_m\le \mu_0$, with $\mu_0$ being the value earlier defined by
Eqs.(\ref{mu0},\ref{tau0}). This condition after a simple algebra
reduces to the inequality
\begin{equation}\label{R*}
R^2\ge
R_*^2=-2\frac{f''(\tau_*)}{f'''(\tau_*)}-\frac{f'(\tau_*)}{f''(\tau_*)}
\end{equation}
for the confinement radius $R$. Here $\tau^*$ is the (unique)
solution of the equation
$\tau_*=-2\frac{f''(\tau_*)}{f'''(\tau_*)}\equiv z(\tau_*)$.
Existence (and uniqueness) of such a solution is ensured for the
short-ranged potentials satisfying our main condition $A(x)>0$.
Indeed, in that case the function $z(\tau)$ satisfies $z(0)>0$ and
$dz/d\tau=1-2A(\tau)/[f'''(\tau)]^2<1$, so that the point of
intersection of the graph of the function $y=z(\tau)$ and the
straight line $y=\tau$ is obviously unique.

As long as $R_{cr}\le R\le R_*$ the position of the maximum on the
AT line is irrelevant as the AT line earlier meets the line
$T_b(\mu)$ at the point $(\mu_0,T_0)$. As we have already seen in
our analysis of the replica-symmetric solution, the line
$T_b(\mu)$ must have at the meeting point the vertical
tangent:$\frac{d}{d\mu}T_B(\mu)|_{\mu_0}=\infty$. The line
$T_b(\mu)$ then continues to the phase with broken replica
symmetry. The corresponding equation is given by solving the
system of equations  Eq.(\ref{Q1RSBa}) and (\ref{m1RSBa}) together
with the condition:
\begin{equation}\label{Rconstrainta}
y_0(Q)-\frac{1}{\mu^2}f'\left[y_0(Q)\right]=R^2\,.
\end{equation}
The resulting transition line $T_b(\mu)$ evidently ends up at zero
temperature $T=0$ at some point $\mu_b$ of the $\mu-$axis. A
typical phase diagram of this sort was presented in fig.1c for the
particular case of $f(x)=e^{-x}$, when the value $R_*$ can be
found analytically as $R_*=\sqrt{3}$, see Eq.(\ref{R*}). Note that
in contrast to the long-ranged case now the line $T_b(\mu)$ is not
strictly vertical inside the glassy phase, although actual
numerical values of its derivative are quite big. Let us also note
that in the $R-$dominated part of the glassy phase (i.e. to the
left of the line $T_b(\mu)$) the values of the parameters
$m,Q,q_0$ at a given temperature $T$ "freeze" to their
$\mu-$independent values which those parameters acquired precisely
at the point of the transition line $T_b(\mu)$ such that $T_b=T$.

It is appropriate to recall that the AT stability line
Eq.(\ref{ATmu}) for a short-range potential meets $T=0$ line at
the point $\mu=\mu_{AT}=[f''(0)]^{1/2}$. Therefore for such
systems a non-trivial transition to the phase with broken replica
symmetry is possible also at zero temperature with decreasing
$\mu$, in contrast to the case of long-ranged potentials. In fact,
it is known that some aspects of zero-temperature behaviour may be
amenable to investigation without resorting to replicas, see
\cite{fluc,FSW}, so it may be useful to provide below a more
detailed picture of the zero-temperature transition within the
replica approach. In performing the zero-temperature limit one
first should realize that $\lim_{T\to 0 }m/T=v<\infty$, so that
the relevant parameters governing the system behaviour in that
case will be $v$ and $Q$. The variational excess (free) energy is
given in terms of these parameters by
\begin{equation}\label{zeroT}
2\Delta F_{T=0}=\mu\,Q +\frac{1}{v}\ln{\left(1-\mu\,v\,Q\right)}+
v\,\left[f(Q)-f(0)-Qf'(0)\right].
\end{equation}
The corresponding stationarity conditions can be reduced after
some algebra to the set of two equations:
\begin{equation}\label{zeroTa}
\frac{\mu^2\,Q}{1-\mu\,v\,Q}=f'(Q)-f'(0),
\end{equation}
\begin{equation}\label{zeroTb}
\frac{1}{v}\ln{\left(1-\mu\,v\,Q\right)}=-\mu\,Q+v\left[f(Q)-f(0)-Qf'(Q)\right],
\end{equation}
which also could be obtained by performing the limit $T\to 0$
directly in Eq.(\ref{Q1RSBa}) and ( \ref{m1RSBa}). A
straightforward calculation shows that close to the transition
point $\mu=\mu_{AT}=[f''(0)]^{1/2}$ the parameter $v$ tends to the
value $v=-\frac{f'''(0)}{2\mu_{AT}^3}$, and the corresponding
excess energy at  zero temperature behaves again cubically in the
vicinity of transition (cf. Eq.(\ref{freeQsmalla}):
\begin{equation}\label{nullT}
\Delta\, F_{T=0}=-\frac{2}{3}\frac{\mu_{AT}^3}{f'''(0)}\left(
\frac{\mu}{\mu_{AT}}-1\right)^3>0\,.
\end{equation}

Let us now consider the case of infinite confinement radius
$R\to\infty$, which was the subject of earlier studies within GVA
\cite{Engel}. In this situation it is interesting to investigate
the behaviour of the system at zero temperature and vanishing
confining potential $\mu\to 0$. Analysis of the system
Eq.(\ref{zeroTa}) then shows that as long as $\mu\to 0$
necessarily $Q\to \infty$ in such a way that $\mu v Q\to 1$,
provided for large $Q$ holds $f(Q)\to 0$ as well as $Qf'(Q)\to 0$.
One then finds after some algebra that the parameter $v$ must
satisfy in this limit the equation
\[
v\,e^{-f(0)v^2}=-\frac{\mu}{f'(0)}\,.
\]
The valid solution $v=v_{\mu}\gg 1$ is given asymptotically by
\begin{equation}\label{vmax}
v_{\mu}^2=-\frac{1}{f(0)}\ln{\left(-\frac{\mu}{f'(0)}\right)}+
O(\ln{|\ln{\mu}|})\,.
\end{equation}
This in turn yields for the parameter $Q$ and for the excess free
energy the following asymptotic expressions:
\begin{equation}\label{vmaxa}
Q\approx\frac{1}{\mu}\frac{1}
{\sqrt{-\frac{1}{f(0)}\ln{\left(-\frac{\mu}{f'(0)}\right)}}},\quad
\Delta{F}\approx-\sqrt{f(0)|\ln{\left(-\frac{\mu}{f'(0)}\right)|}}\,.
\end{equation}
Recalling that for zero temperature $q_d=Q-\frac{1}{\mu^2}f'(Q)$,
hence $Q\to \infty$ implies $q_d \approx Q$, we indeed see that
Eq.(\ref{vmaxa}) agrees with one obtained in the framework of
GVA\cite{Engel} for the mean displacement parameter
$q_d=\frac{1}{N}\left\langle {\bf x}^2\right \rangle$. One also
can perform similar calculations in the opposite limiting case
$\mu=0$ and $R\to \infty$, the result for $T=0$ being given by the
same asymptotic formulae Eq.(\ref{vmaxa}) with the only
replacement $\mu\to 1/R^2$.


Returning to considerations of general finite values of the
parameter $R$, we see that the $\mu-$dominated regime described at
$T=0$ by the equations Eq.(\ref{zeroTa}) can exist as long as
\begin{equation}\label{zeroTc}
Q-\frac{1}{\mu^2}f'(Q)\le R^2\,,
\end{equation}
where we took into account the zero-temperature limit of
Eq.(\ref{q01RSBa}). With decreasing $\mu$ the latter condition is
first violated at some value $\mu=\mu_b$ which can be found
numerically by solving the system of three equations:
Eq.(\ref{zeroTa},\ref{zeroTb}) and (\ref{zeroTc}). For lower
values of $\mu$, i.e. for $\mu<\mu_b$, the parameters $Q$ and $v$
will "freeze" at their boundary values $v=v(\mu_b)$ and
$Q=Q(\mu_b)$.

The shape of the phase diagram described above as typical for
$R<R_*$, see Eq.(\ref{R*}), experiences essential modifications as
long as $R$ exceeds $R_*$. In the latter case the AT Temperature
line $T_{AT}(\mu)$ starts to decrease with decreasing $\mu$ to the
left of the maximum: $\mu<\mu_m$. We shall see that the correct
form of the phase diagram in this situation requires considering
the transition line to the phase with broken replica symmetry
given by the condition $m\to 1$.

To investigate such a possibility we expand the excess free energy
Eq.(\ref{free1RSBb}) in powers of $p=1-m$ as
\begin{equation}\label{freem}
\Delta F=A(Q)\,p+B(Q)\,p^2+C(Q)\,p^3+\ldots,
\end{equation}
Extremising with respect to $p$ gives the condition
$A(Q)+2p\,B(Q)+3p^2\,C(Q)+\ldots =0$ so that the equilibrium value
of $p$ for small $p$ is approximately given by $p_m\approx
-A(Q)/2B(Q)+\ldots$. The excess free energy in the phase with
broken replica symmetry is then given by $\Delta F\approx
A(Q)\,p_m+B(Q)p_m^2=-A(Q)^2/4B(Q)\ge 0$, which implies $B(q)\le
0$. As $p_m\ge 0$, we conclude $A(Q)\ge 0$. It is evident also
that the line at which $p_m\to 0$ implies the condition $A(Q)=0$.
Moreover, extremum conditions of Eq.(\ref{freem}) with respect to
$Q$ implies in the limit $p_m\to 0$ another condition $A'(Q)=0$.
This consideration shows that the line at which $m\to 1$ within
the phase with broken replica symmetry is given by the system of
equations:
\begin{equation}\label{A}
A(Q)=0,\quad \frac{d}{dQ}A(Q)=0,
\end{equation}
where explicit expression for $A(Q)$ can be found from the
condition $A(Q)=\frac{\partial \Delta F}{\partial p}|_{p=0}$. This
gives, up to an unessential factor
\begin{equation}\label{A1}
A(Q)=\frac{\mu}{T}Q+\ln{\left(1-\frac{\mu}{T}Q\right)}+\frac{1}{T^2}
\left[f\left(\frac{T}{\mu}-Q\right)-f\left(\frac{T}{\mu}\right)+Qf'\left(\frac{T}{\mu}\right)\right]
\end{equation}
and therefore
\begin{equation}\label{A2}
A'(Q)=-\frac{\mu^2}{T^2}\frac{Q}{1-\frac{\mu}{T}Q}-\frac{1}{T^2}
\left[f'\left(\frac{T}{\mu}-Q\right)-f'\left(\frac{T}{\mu}\right)\right]\,.
\end{equation}
Substituting these expressions to equations Eq.(\ref{A}) we see
that the resulting conditions are indeed equivalent to the pair of
equations (\ref{Q1RSBa}) and (\ref{m1RSBa}), provided $m\to 1$. We
will be looking for its non-vanishing solution $Q=Q_*>0$.

The equation Eq.(\ref{A2}) for $Q\to 0$ reads $A'(Q)\approx
\frac{Q}{T^2}\left[-\mu^2+f''(T/\mu)\right]$, hence $A'(Q)$ is
positive just below the AT instability line Eq.(\ref{ATmu}). We
may expect this sign to change in the vicinity of the point of
maximum $(\mu_m,T_m)$ from Eq.(\ref{Tc}), as we know that the line
$m=1$ passes through that point. In the vicinity of AT line the
parameter $Q$ is small, and we can expand
\begin{equation}\label{A3}
A(Q)=-\frac{Q^2}{2T^2}\left[\mu^2-f''\left(\frac{T_m}{\mu_m}\right)\right]-\frac{1}{3}\frac{Q^3}{T^2}\left[\frac{\mu^3}{T}+\frac{1}{2}f'''\left(\frac{T}{\mu}\right)\right]
+\frac{Q^4}{4T^2}\left[\frac{1}{6}f''''(T/\mu)-\frac{\mu^4}{T^2}\right]+\ldots
\end{equation}
One indeed finds from this expression a non-vanishing solution
$Q_*>0$ of the system Eq.(\ref{A}) for $T<T_m,\mu<\mu_m$ close to
the point of maximum to be
\begin{equation}\label{A4}
Q_*=\frac{2}{3}\frac{\frac{\mu^3}{T}+\frac{1}{2}f'''\left(\frac{T}{\mu}\right)}{\frac{1}{6}f''''\left(\frac{T_m}{\mu_m}\right)-\frac{\mu^4}{T^2}},\quad
T-T_m\ll T_m,\,\mu-\mu_m\ll \mu_m
\end{equation}
Substituting this back to Eq.(\ref{A1}) we get an explicit
relation between $T$ and $\mu$ defining the line $T_1(\mu)$ along
which a solution with broken replica symmetry $Q\ne 0$ appears
first with $m=1$ close to $(\mu_m,T_m)$:
\begin{equation}\label{A5}
\left[\mu^2-f''\left(\frac{T_m}{\mu_m}\right)\right]\left[\frac{1}{6}f''''\left(\frac{T_m}{\mu_m}\right)-\frac{\mu^4}{T_1^2}\right]+
\frac{2}{9}\left[\frac{\mu^3}{T_1}+\frac{1}{2}f'''\left(\frac{T_1}{\mu}\right)\right]^2=0,
\end{equation}
Expanding in powers of $\tau_1=(T_m-T_1)/T_m\ll
1,\,\,\delta=(\mu_m-\mu)/\mu_m\ll 1$ one can find to the leading
order the equation for this line to be
\begin{equation}\label{A6}
\tau_1=\frac{\delta^2}{2}\left[1-\frac{T_m^2}{6\mu_m^4}f''''\left(\frac{T_m}{\mu_m}\right)\right],\quad
\delta\le 0,
\end{equation}
which shows that the transition line $T_1(\mu)$ when approaching
the point $(\mu_m,T_m)$ from the left has in that point its
maximum. At the point of maximum it therefore smoothly meets the
AT line which in the same approximation is described by
Eq.(\ref{A7}), with only second derivative experiencing a jump.

For lower values of $\mu$ away from $\mu_c$ the temperature
$T_1(\mu)$ decreases down to a point where that line meets the
boundary line $T_b(\mu)$, and the system transits to $R-$dominated
regime. The point of intersection can be found only numerically. For
even lower values of $\mu$ the transition temperature $T_1(\mu)$
freezes to its value at the intersection point, see fig.4. In the
limiting case of infinite confinement radius $R=\infty$ the
$\mu-$dominated regime covers the whole phase diagram, and it makes
sense to investigate the asymptotic behaviour of the line $T_1(\mu)$
for $\mu\to 0$. Anticipating the behaviour $T_1|_{\mu\to 0}\to 0$ in
such a way that $T_1 >> \mu$ one can show that the corresponding
non-vanishing solution $Q_*>0$ of the system Eq.(\ref{A}) must be
sought close to its maximal possible value $Q_*=Q_*^{max}=T/\mu$.
Using $f(\infty)=f'(\infty)=0$ one finds after some  algebra the
corresponding asymptotic expression for the transition line
$T=T_1(\mu)$:
\begin{equation}\label{T1mu}
\mu=-\frac{f'(0)}{T}\exp{-\frac{f(0)}{T^2}},\quad T\to 0\,.
\end{equation}

\begin{figure}[h!]
\centering \hspace*{0.2cm} \includegraphics[width=8cm]{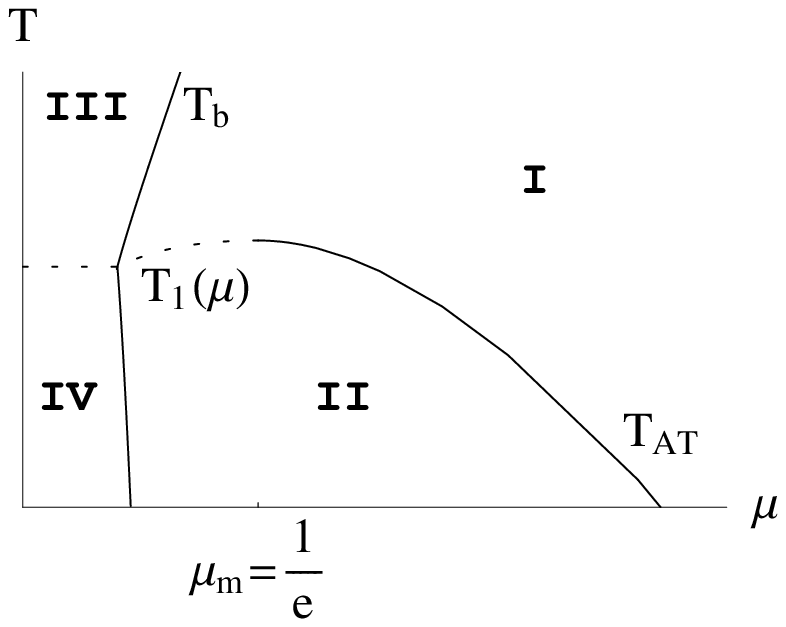}
\caption*{(a)}
\includegraphics[width=8.0cm]{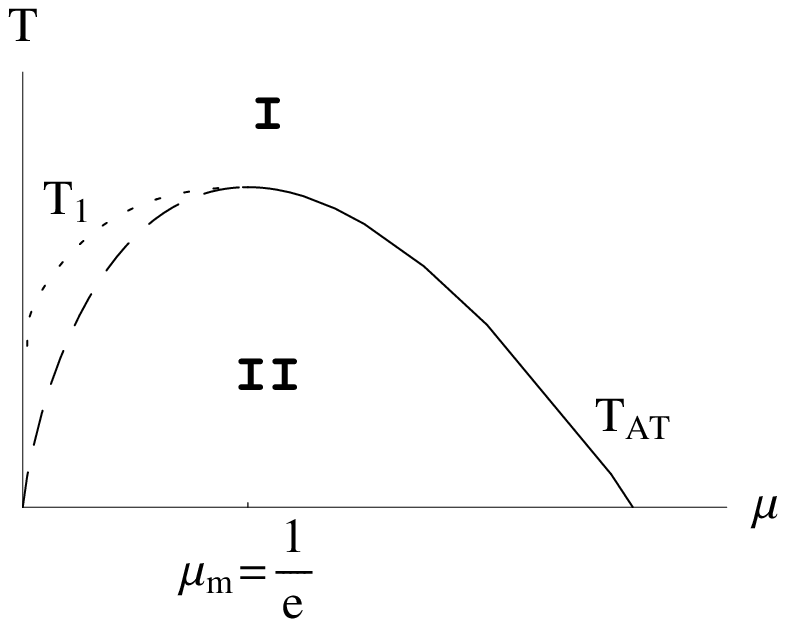}
\caption*{(b)}
\includegraphics[width=8.0cm]{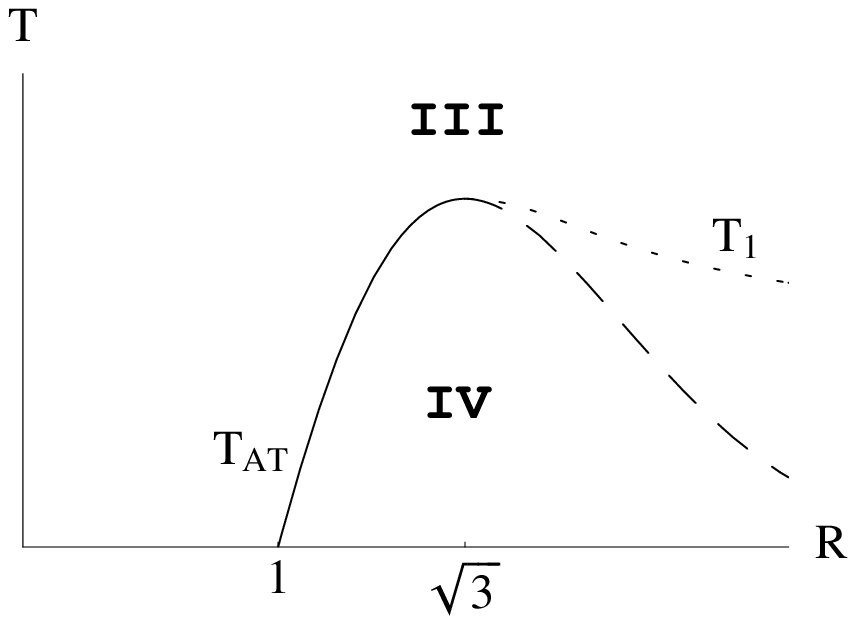}
\caption*{(c)} \caption*{ Figure 4: The phase diagrams for the
potential with short-range correlations, $f(x)=e^{-x}$. In the case
(a) the radius of the box is $R=\sqrt{5}>R_{*}$, in the case (b)
$R=\infty$ and the case (c) corresponds to $\mu=0$. The dotted
curves in (a)-(c) represents the transition line $T_1(\mu)$ found
from the condition $m\to 1$ and the broken curve in (B) and (c)
represents the part of $T_{AT}$ line below $T_1$. The notation for
phases are the same as in fig. 1.}

\centering
\end{figure}

For completeness, we present in fig.4 also typical phase diagrams
for systems with short-range correlations in two limiting cases: in
$(T,\mu)$ plane for $R=\infty$ and in $(T,R)$ plane for $\mu=0$.
Recall, that in the latter case the de Almeida-Thouless temperature
for our choice of the correlation function is explicitly given by
$T_{AT}=T_0(R)=(R^2-1)e^{-\frac{1}{2}(R^2-1)}$.

{\bf Acknowledgements}. This research was accomplished during the
first author's stay as a Bessel awardee at the Institute of
Theoretical Physics, University of Cologne, Germany. Y.V.F. is
grateful to M. Zirnbauer for the kind hospitality extended to him
during his months in Cologne and for stimulating interest in this
work. The Humboldt Foundation is acknowledged for the financial
support of that visit. Both authors are grateful to W. Wieczorek for
his help with preparing figures for this article. The research in
Nottingham was supported by EPSRC grant EP/C515056/1 "Random
Matrices and Polynomials: a tool to understand complexity", and in
Duisburg by the SFB/TR 12 der DFG .

\label{l}
\appendix

\section{Appendix: Parisi matrix, its eigenvalues and evaluation of traces
in the replica limit.}\label{AppendixB}

We start with describing the well-known structure of the $n\times
n$ matrix $Q$ in the Parisi parametrisation. At the beginning we
set $n$ diagonal entries $q_{\alpha\alpha}$ all to the same value
$q_{\alpha\alpha}=0$. This value will be maintained at every but
last step of the recursion. The off-diagonal part of the matrix
$Q$ in the Parisi scheme is built recursively as follows. At the
first step we single out from the $n\times n$ matrix $Q$ the chain
consisting of $n/m_{1}$ blocks of the size $m_1\le n$, each
situated on the main diagonal. All off-diagonal entries
$q_{\alpha\beta},\,\alpha\ne \beta$ inside those blocks are filled
in with the same value $q_{\alpha\beta}=q_1\le 0$, whereas all the
remaining $n^2(1-1/m_1)$ entries of the matrix $Q$ are set to the
value $0<q_0\le q_1$. The latter entries remain from now on intact
to the end of the procedure, whereas some entries inside the
diagonal $m_1\times m_1$ blocks will be subject to a further
modification. At the next step of iteration in each of those
diagonal blocks of the size $m_1$ we single out the chain of
$m_2/m_1$ smaller blocks of the size $m_2\le m_1$, each situated
on the main diagonal. All off-diagonal entries
$q_{\alpha\beta},\,\alpha\ne \beta$ inside those sub-blocks are
filled in with the same value $q_{\alpha\beta}=q_2\ge q_1$,
whereas all the remaining entries of the matrix $Q$ hold their old
values. At the next step only some entries inside diagonal blocks
of the size $m_2$ will be modified., etc. Iterating this procedure
step by step one obtains after $k$ steps a hierarchically built
structure characterized by the sequence of integers
\begin{equation}\label{parisiseq}
n=m_0\ge m_1\ge m_2\ge \ldots\ge m_k\ge m_{k+1}=1
\end{equation}
and the values placed in the diagonal blocks of the $Q$ matrix
satisfying:
\begin{equation}\label{parisiseq1}
0<q_0\le q_1\le q_2\le \ldots\le q_k
\end{equation}
Finally, we complete the procedure by filling in the $n$ diagonal
entries $q_{\alpha\alpha}$ of the matrix $Q$ with one and the same
value $q_{\alpha\alpha}=q_d\ge q_k$.

For the subsequent analysis we need the eigenvalues of the Parisi
matrix $Q$. Those can be found easily together with the
corresponding eigenvectors built according to a recursive
procedure which uses the sequence Eq.(\ref{parisiseq}). It is
convenient to visualize eigenvectors as being "strings" of $n$
boxes numbered from $1$ to $n$, with $l^{th}$ component being a
content of the box number $l$.

At the first step $i=1$ we choose the eigenvector to have all $n$
boxes filled with the same content equal to unity. The
corresponding eigenvalue is non-degenerate and equal to
\begin{equation}\label{parisieigen1}
\lambda_1=q_d+q_k(m_k-1)+q_{k-1}(m_{k-1}-m_k)+
\ldots+q_1(m_1-m_2)+q_0(m_0-m_1)
\end{equation}
Now, at the subsequent steps $i=2,3,\ldots,k+2$ one builds
eigenvectors by the following procedure. The string of $n$ boxes
of an eigenvector belonging to $i^{th}$ family are subdivided into
$n/m_{i-1}$ substrings of the length $m_{i-1}$, and numbered
accordingly by the index $j=1,2,\ldots, n/m_{i-1}$. All $m_{i-1}$
boxes of the first substring $j=1$ are filled invariably with all
components equal to $1$. Next we fill $m_{i-1}$ boxes in one (and
only one) of the remaining $\frac{n}{m_{i-1}}-1$ substrings with
all components equal to $-1$. In doing so we however impose a
constraint that the substrings with the indices $j$ given by
$j=1+l\frac{m_{i-2}}{m_{i-1}}$ should be excluded from the
procedure, with $l$ being any integer satisfying $1\le l\le
\frac{n}{m_{i-2}}-1$. After the choice of a particular substring
is made, we fill all $n-2m_{i-1}$ boxes of the remaining
substrings with identically zero components. It is easy to see
that all $d_i=n/m_{i-1}-n/m_{i-2}$ different eigenvectors of
$i^{th}$ family built in such a way correspond to one and the same
$d_i-$degenerate eigenvalue
\begin{equation}\label{parisieigeni} \lambda_i=q_d+q_k(m_k-1)+q_{k-1}(m_{k-1}-m_k)+
\ldots+q_{i-1}(m_{i-1}-m_i)-q_{i-2}(m_{i-1})
\end{equation}
In this way we find all $n$ possible eigenvalues, the last being
equal to
\begin{equation}\label{parisieigenk} \lambda_{k+2}=q_d-q_k\,
m_{k+1}\equiv q_d-q_k.
\end{equation}
The completeness of the procedure follows from the fact that sum
of all the degeneracies $d_i$ is equal to
\[
1+\left(\frac{n}{m_1}-1\right)+\left(\frac{n}{m_2}-\frac{n}{m_1}\right)+\ldots+
\left(\frac{n}{m_{k+1}}-\frac{n}{m_k}\right)=n
\]
Note that all the found eigenvalues are positive due to
inequalities Eq.(\ref{parisiseq1}) between various $q_i$, which is
required by the positive definiteness of the matrix $Q$. Note also
that all eigenvectors built in this way are obviously linearly
independent, although the eigenvectors belonging to the same
family are not orthogonal. The latter fact however does not have
any bearing for our considerations.

To facilitate the subsequent treatment it is convenient to
introduce the following (generalized) function of the variable
$q$:
\begin{equation}\label{xstep}
x(q)=n+\sum_{l=0}^k (m_{l+1}-m_l)\,\theta(q-q_l)
\end{equation}
where we use the notation $\theta(z)$ for the Heaviside step
function: $\theta(z)=1$ for  $z>0$ and zero otherwise. In view of
the inequalities Eq.(\ref{parisiseq},\ref{parisiseq1}) the
function $x(q)$ is piecewise-constant non-increasing, and changes
between $n$ and $1$ as follows:
\begin{equation}\label{xstep1}
x(q<q_0)=m_0\equiv
n,\,\,x(q_0<q<q_1)=m_1,\,\ldots,\,x(q_{k-1}<q<q_k)=m_k,\,x(q>q_k)=m_{k+1}\equiv
1
\end{equation}
Comparison of this form with Eq.(\ref{xstep}) makes evident the
validity of a useful inversion formula:
\begin{equation}\label{xstep2}
\frac{1}{x(q)}=\frac{1}{n}+\sum_{l=0}^k
\left(\frac{1}{m_{l+1}}-\frac{1}{m_l}\right)\,\theta(q-q_l)
\end{equation}
which will be exploited by us shortly.

 As observed by Crisanti and Sommers\cite{sph1} one
can represent the eigenvalues Eq.(\ref{parisieigeni}) of the
Parisi matrix in a compact form via the following remarkable
identities:
\begin{equation}\label{criso}
\lambda_{1}=\int_{0}^{q_d}x(q)\,dq=nq_0+\int_{q_0}^{q_d}x(q)\,dq,\quad\lambda_{i+2}=\int_{q_i}^{q_d}x(q)\,dq,\quad
i=0,1,\ldots,k
\end{equation}
As a consequence, these relations imply for any analytic function
$g(x)$ the identity
\begin{equation}\label{criso1}
\frac{1}{n}Tr\left[
g(Q)\right]=\frac{1}{n}\sum_{i=1}^{k+2}g(\lambda_{i})\,d_i
=\frac{1}{n}g\left(nq_0+\int_{q_0}^{q_d}x(q)\,dq\right)+
\sum_{l=0}^{k}\left(\frac{1}{m_{l+1}}-\frac{1}{m_{l}}\right)g\left(\int_{q_l}^{q_d}x(q)\,dq\right).
\end{equation}
Next one observes that taking the derivative of the generalized
function from Eq.(\ref{xstep2}) produces
\begin{equation}\label{xstep3}
\frac{d}{dq}\left[\frac{1}{x(q)}\right]=\sum_{l=0}^k
\left(\frac{1}{m_{l+1}}-\frac{1}{m_l}\right)\,\delta(q-q_l).
\end{equation}
This fact allows one to rewrite the sum in Eq.(\ref{criso1}) in
terms of an integral, yielding
\[
\frac{1}{n}Tr\left[ g(Q)\right]
=\frac{1}{n}g\left(nq_0+\int_{q_0}^{q_d}x(q)\,dq\right)+
\int_{q_0-0}^{q_k+0}g\left(\int_{q}^{q_d}x(\tilde{q})\,d\tilde{q}\right)\,\frac{d}{dq}\left[\frac{1}{x(q)}\right]\,dq,
\]
where the short-hand notation $q\pm 0$ designates the limit from
below/above. Further performing integration by parts, and using
$x(q>q_k)=1,\,x(q<q_0)=n$, we finally arrive at
\begin{eqnarray}\label{criso11}
&& \frac{1}{n}Tr\left[ g(Q)\right]
=\frac{1}{n}\left[g\left(nq_0+\int_{q_0}^{q_d}x(q)\,dq\right)-g\left(\int_{q_0}^{q_d}x(q)\,dq\right)\right]
\\ \nonumber
&+& g(q_d-q_k)+\int_{q_0}^{q_k}g'
\left(\int_{q}^{q_d}x(\tilde{q})\,d\tilde{q}\right)dq.
\end{eqnarray}
We are actually interested in the replica limit $n\to 0$.
According to the Parisi prescription in such a limit the
inequality Eq.(\ref{parisiseq}) should be reversed:
\begin{equation}\label{parisiseq2a}
n=0\le m_1\le m_2\le \ldots\le m_k\le m_{k+1}=1
\end{equation}
and the function $x(q)$ is now transformed to a non-decreasing
function of the variable $q$ in the interval $q_0\le q \le q_k$,
and satisfying outside that interval the following properties
\begin{equation}\label{outsidea}
x(q<q_0)=0,\quad \mbox {and}\quad  x(q>q_k)=1.
\end{equation}
In general,such a function also depends on the increasing sequence
of $k$ parameters $m_l$ described in Eq.(\ref{parisiseq2}) .

The form of Eq.(\ref{criso11}) makes it easy to perform the limit
$n\to 0$ explicitly, and to obtain after exploitation of
Eq.(\ref{outside}) an important identity Eq.(\ref{traces}) helping
to evaluate the traces in the replica limit. Finally, let us
mention the existence of an efficient method of the "replica
Fourier transform" allowing one to diagonalise (and otherwise
work) with much more general types of hierarchical matrices, see
\cite{CDT,DCT} for more detail.

\section{Appendix: Stability analysis of the replica solutions.}
\label{AppendixA}
\subsection{General relations.}
 In this part we are going to derive
the general stability equations in the $\mu-$ dominated regime, and
then indicate modifications arising in the $R-$ dominated case.

In general, the stability analysis amounts to expanding the function
$\Phi(Q)$ in Eq.(14) around the extremum point up to the second
order in deviations: $\Phi=\Phi_{SP}+ \delta \Phi + {1\over
2}{\delta}^2 \Phi$. In the $\mu-$ dominated regime the stationary
point is inside the integration region and the stationarity
condition amounts to $\delta \Phi =0$ yielding the system (16)-(17).
The term ${\delta}^2 \Phi$ is a quadratic form in independent
fluctuation variables $\delta q_{ab}, a\le b$ and can be generally
written as ${\delta}^2 \Phi=\sum_{(ab),(cd)}\delta q_{(ab)}
G_{(ab),(cd)}\,\delta q _{(cd)}$. As usual the stable extremum
corresponds to the positive definite quadratic form, and along the
critical line the quadratic form becomes semi-definite. Checking
positive definiteness of  ${\delta}^2 \Phi$ amounts to finding the
(generalized) eigenvalues $\Lambda$ of the matrix
$G_{(ab),(cd)}=\frac{\partial^2\Phi}{\partial q_{ab}\partial
q_{cd}}$, i.e. solving the equations
\begin{equation} \label{stab1}
\sum_{(cd),c\le d} G_{(ab),(cd)}\,\eta_{(cd)}=\Lambda
\sum_{(cd),c\le d} C_{(ab),(cd)}\,\eta_{(cd)},\quad a\le b,
\end{equation}
Here the notation $\eta_{ab}, \, a\le b$ is used for $n(n+1)/2$
components of a (generalized) eigenvector $\eta$ and $C>0$ can be
any real symmetric positive definite matrix used to define a
suitable scalar product $\left(\delta q^{(A)},\delta
q^{(B)}\right)_C$ in the space of fluctuation vectors $\delta q$,
that is
\begin{equation}\label{scalar}
\left(\delta q^{(A)},\delta q^{(B)}\right)_C= \sum_{(ab),(cd)}
\delta q^{(A)}_{(ab)}C_{(ab),(cd)}\delta q^{(B)}_{(cd)}.
\end{equation}
Indeed, for any choice of $C$ all generalized eigenvalues are real
due to the symmetry properties of the matrix $G$ , and the
eigenvectors $\eta^{(i)}$ and $\eta^{(j)}$ corresponding to
different eigenvalues $\Lambda_i\ne \Lambda_j$ are orthogonal with
respect to the chosen scalar product:
$\left(\eta^{(i)},\eta^{(j)}\right)_C=0$, hence linearly independent
and form a basis. Expanding the fluctuations in this basis as
$\delta q=\sum_i p_i\eta^{(i)}$ one then finds generically
${\delta}^2 \Phi=\sum_{j} \Lambda_i
p_i^2\left(\eta^{(i)},\eta^{(i)}\right)_C$, and the positive
definiteness amounts to the condition $\Lambda_i>0$ for all $i$. The
instability occurs when some of the eigenvalues vanish:
$\Lambda_i=0$, and it is easy to see that the condition for the
instability is independent of the choice of the matrix $C$ in the
definition of the scalar product Eq.(\ref{scalar}). In fact, in our
analysis we find it convenient to introduce a formal convention
$\eta_{(ab)}=\eta_{(ba)}$ for eigenvector components, which makes it
natural to think of the eigenvectors $\eta$ as being real symmetric
matrices. Accordingly, we find it convenient to define the scalar
product in that vector space for any two such eigenvectors
$\eta^{(A)}$ and $ \eta^{(B)}$ as
\begin{equation}\label{scalar1}
\left(\eta^{(A)},\eta^{(B)}\right)=\mbox{Tr}
\left[\eta^{(A)}\eta^{(B)}\right]
\end{equation}
This simply corresponds to choosing the matrix $C$ in
Eq.(\ref{scalar}) to be diagonal, with the diagonal entries given by
$C_{(ab),(ab)}=(2-\delta_{ab})$.

Our first task is to determine explicitly the entries of the
stability matrix $G$ for our problem.
 The structure of the replica free energy functional
Eq.(\ref{repham1}) suggests to represent the matrix $G$ as a sum of
two terms $G=G_I+G_{II}$ such that:
\begin{equation} \label{stab2}
\left( G_I\right)_{(ab),(cd)}=-\frac{T}{2}\frac{\partial^2}{\partial
q_{ab}\partial q_{cd}}\ln{\det{Q}},\quad \left(
G_{II}\right)_{(ab),(cd)}=-\frac{1}{2T}\frac{\partial^2}{\partial
q_{ab}\partial q_{cd}}\,\sum_{(a\ne b)}f(D_{ab}),
\end{equation}
where the last summation is assumed to go over {\it all} possible
pairs $(ab)$ with $a\ne b$, and we introduced a short-hand notation
$D_{ab}=\frac{1}{2}(q_{aa}+q_{bb}-2q_{ab})$. A straightforward
application of the Wick theorem shows that:
\begin{equation} \label{stab3a}
\left( G_I\right)_{(a\ne b),(c \ne
d)}=T\left[\left(Q^{-1}\right)_{ac} \left(Q^{-1}\right)_{bd}+
\left(Q^{-1}\right)_{ad} \left(Q^{-1}\right)_{bc}\right],\,\, \left(
G_I\right)_{(aa),(cc)}=
\frac{T}{2}\left[\left(Q^{-1}\right)_{ad}\right]^2,\end{equation}

\begin{equation} \label{stab3b}
\left( G_I\right)_{(a\ne b),(cc)}=T\left(Q^{-1}\right)_{ac}
\left(Q^{-1}\right)_{bc},\,\, \left(
G_I\right)_{(aa),(bc)}=T\left(Q^{-1}\right)_{ab}\left(Q^{-1}\right)_{ac}.
\end{equation}
In view of the convention $\eta_{(ab)}=\eta_{(ba)}$ for eigenvector
components, we have:
\begin{equation} \label{stab4}
\sum_{(cd),c\le d} \left( G_I\right)_{(ab),(cd)}\,\eta_{(cd)}=
\frac{T}{1+\delta_{ab}}
\sum_{(cd)}\left(Q^{-1}\right)_{ac}\left(Q^{-1}\right)_{bd}\eta_{(cd)},
\end{equation}
where $\delta_{ab}$ stands for the Kronecker symbol, and the
summation in the right-hand side goes over all pairs of indices
without restrictions.

Further simple differentiation gives for the entries of $G_{II}$
with $a\ne b$ the expressions:
\begin{equation} \label{stab5a}
\left( G_{II}\right)_{(a\ne b),(a\ne b)}=-\frac{1}{T}
\,f''(D_{ab}),\,\,\left( G_{II}\right)_{(a\ne b),(aa)}=\left(
G_{II}\right)_{(a\ne b),(bb)}=\frac{1}{2T} \,f''(D_{ab}),\,\,
\end{equation}
and $\left( G_{II}\right)_{(a\ne b),(cd)}=0$ for all other choices
of $(cd)$. If however $a=b$, we similarly find
\begin{equation} \label{stab5b}
\left( G_{II}\right)_{(aa),(aa)}=-\frac{1}{4T}\sum_{c,\,c\ne d}
\,f''(D_{ac}),\,\,\left( G_{II}\right)_{(aa),(ab)}=\frac{1}{2T}
\,f''(D_{ab}),\,\, \left( G_{II}\right)_{(aa),(bb)}=-\frac{1}{4T}
\,f''(D_{ab}),
\end{equation}
and $\left( G_{II}\right)_{(aa),(cd)}=0$ for all other choices of
$(cd)$. We therefore see that
\begin{equation} \label{stab6a}
\sum_{(cd),c\le d} \left( G_{II}\right)_{(ab),(cd)}\,\eta_{(cd)}=
\frac{1}{2T}f''(D_{ab})\left[\eta_{(aa)}+\eta_{(bb)}-2\eta_{(ab)}\right],\quad
a\ne b,
\end{equation}
and for $a=b$
\begin{equation} \label{stab6b}
\sum_{(cd),c\le d} \left( G_{II}\right)_{(aa),(cd)}\,\eta_{(cd)}=
-\frac{1}{4T}\sum_{c,\,c\ne a}
f''(D_{ac})\left[\eta_{(aa)}+\eta_{(cc)}-2\eta_{(ac)}\right].
\end{equation}
Combining all these expressions and definitions we see that the
system of equations Eq.(\ref{stab1}) for eigenvector component
$\eta_{(ab)}$ and eigenvalues $\Lambda$ of the stability matrix $G$
in the $\mu-$dominated regime takes the form:
\begin{equation}\label{stab7a}
T\,\sum_{(cd)}\left(Q^{-1}\right)_{ac}\left(Q^{-1}\right)_{bd}\eta_{(cd)}+
\frac{1}{2T}f''(D_{ab})\left[\eta_{(aa)}+\eta_{(bb)}-2\eta_{(ab)}\right]
=2\Lambda\,\eta_{(ab)},\quad a\ne b
\end{equation}
and for the eigenvector components with $a=b$:
\begin{equation}\label{stab7b}
\frac{T}{2}\,\sum_{(cd)}\left(Q^{-1}\right)_{ac}\left(Q^{-1}\right)_{ad}\eta_{(cd)}
-\frac{1}{4T}\sum_{c}f''(D_{ac})\left[\eta_{(aa)}+\eta_{(cc)}-2\eta_{(ac)}\right]
=\Lambda\,\eta_{(aa)}.
\end{equation}
Assuming $T>0$ and introducing the notation $\Lambda^*=2\Lambda/T$,
we can rewrite the above pair in the unified form:
\begin{equation}\label{unified}
 \sum_{cd} \left ( Q^{-1} \right )_{ac}
 \eta_{(cd)} \left ( Q^{-1} \right )_{db} + {1 \over T^2}
 f''(D_{ab}) (\delta D_{ab})-{1 \over T^2}\delta_{ab}
 \sum_{c}f''(D_{ac}) (\delta D_{ac}) = \Lambda^* \eta_{(ab)}
\label{eigenequ}
\end{equation}
where we used the notation
\begin{equation}\label{DD}
\delta D_{ab}=
\frac{1}{2}\left[\eta_{(aa)}+\eta_{(bb)}-2\eta_{(ab)}\right].
\end{equation}

Let us now indicate minor changes required in the similar procedure
for $R-$dominated regime. First of all, in that regime the diagonal
entries $q_{aa}$ are fixed to the boundary value $q_{aa}=R^2,\forall
a$. The stability for the linear deviations $\delta \Phi>0$ implies
$\partial \Phi/\partial q_{aa}<0$ at fixed $q_{ab}$ on the boundary,
which amounts to the inequality
\begin{equation}\label{linstab}
\mu/T<\sum_b (Q^{-1})_{ab}\,.
\end{equation}
 On the other hand, in the
$\mu-$dominated regime the last inequality is replaced by equality.
Hence, it is easy to see that the inequality (\ref{linstab}) means
nothing but $\mu<\mu_0$ with $T=T_b(\mu_0)$.

As to the quadratic form $\delta^2 \Phi$, the entries of the matrix
$G$ involve in this regime only pairs of indices with $(a\ne b)$ and
$(c\ne d)$. The explicit expressions for those entries are given by
formally the same equations as in the $\mu-$dominated regime. The
eigenvectors $\eta$ must now have only components $\eta_{(a<b)}$. It
is again convenient to introduce the convention
$\eta_{(ab)}=\eta_{(ba)}, a\ne b$ for those components, together
with $\eta_{(aa)}=0,\forall\, a$. The corresponding eigen-equation
then amounts to
\begin{equation}\label{stab8}
\sum_{(cd)}\left(Q^{-1}\right)_{ac}\left(Q^{-1}\right)_{bd}\eta_{(cd)}-
\frac{1}{T^2}f''\left(R^2-q_{ab}\right)\,\eta_{(ab)}
=\Lambda^*\,\eta_{(ab)},\quad a\ne b.
\end{equation}

Now we proceed to the analysis of the main equations
Eq.(\ref{stab7a},\ref{stab7b}) and (\ref{stab8}) in various cases.

\subsection{Stability of the replica symmetric solution.}
 The entries of the matrix $Q$ in this case are given by
$q_{ab}=q_0+(q_d-q_0)\delta_{ab}$, so its inverse $Q^{-1}$ has the
same form $(Q^{-1})_{ab}=p_0+(p_d-p_0)\delta_{ab}$, with $p_0$ and
$p_d$ defined in Eqs.(\ref{invsym1},\ref{invsym2}). Taking this fact
reduces the equation Eqs.(\ref{stab7a}) for $\quad a\ne b$ to the
form (see Eq.(\ref{DD})):
\begin{equation}\label{stab7aa}
p_0^2\,\sum_{(cd)}\eta_{(cd)}+p_0(p_d-p_0)\,\left[\sum_{d}\eta_{(ad)}+
\sum_{d}\eta_{(bd)}\right]+(p_d-p_0)^2\,\eta_{(ab)}+\frac{1}{T^2}
f''(D)\delta D_{ab} =\Lambda^*\,\eta_{(ab)}
\end{equation}
where $D=q_d-q_0$. The equation Eq.(\ref{stab7b}) for $a=b$ is
similarly reduced to
\begin{equation}\label{stab7bb}
p_0^2\,\sum_{(cd)}\eta_{(cd)}+2p_0(p_d-p_0)\,\sum_{d}\eta_{(ad)}+
(p_d-p_0)^2\,\eta_{(aa)}-\frac{1}{T^2} f''(D)\sum_{c}\delta D_{ac}
=\Lambda^*\,\eta_{(aa)},\quad \forall\, a
\end{equation}

Now we can follow faithfully the lines of the classical work by De
Almeida and Thouless \cite{AT} and to provide an explicit
construction of the eigenvectors with components $\eta_{(ab)}$.

The first ("longitudinal") family consists of those eigenvectors
which have the same form as the replica-symmetric $Q$ matrices
themselves: $\eta^{(I)}_{(ab)}=\beta+(\alpha-\beta)\delta_{ab}$.
Then $\sum_{d}\eta^{(I)}_{(ad)}=\alpha+\beta\,(n-1)$, and
Eqs.(\ref{stab7aa},\ref{stab7bb}) is reduced in the replica limit
$n\to 0$ to a system of two equations
\begin{equation}\label{long1}
2p_0(p_d-p_0)(\alpha-\beta)+(p_d-p_0)^2\,\beta+\frac{1}{T^2}
f''(D)(\alpha-\beta)=\Lambda^*\,\beta
\end{equation}
\begin{equation}\label{long2}
2p_0(p_d-p_0)(\alpha-\beta)+(p_d-p_0)^2\,\alpha+\frac{1}{T^2}
f''(D)(\alpha-\beta)=\Lambda^*\,\alpha
\end{equation}

One can find the corresponding two eigenvalues easily, but for our
goals it is enough to notice that for any values of the system
parameters the eigenvalue $\Lambda^*$ can not vanish as
$\Lambda^*=0$ immediately implies $\beta=\alpha=0$. Hence,
fluctuations in the longitudinal family cannot induce instability of
the replica-symmetric solution.

Second family $\eta^{(II)}$ of the eigenvectors can be constructed
as those characterized by one particular replica index singled out
among all $n$. As all these choices are equivalent up to permutation
of replica indices, we can take for definiteness $a=1$. At this
special choice of the replica index we have explicitly
\begin{equation}\label{fam2a}
\eta^{(II)}_{(11)}=\omega,\,\,\eta^{(II)}_{(1b)}=\eta^{(II)}_{(b1)}=\tau,\,\forall\,
b\ne 1,\, \mbox{and}\,
\eta^{II}_{(ab)}=\mu+\delta_{ab}(\nu-\mu),\,\forall\, a\ne
1\,;\forall\,b\ne 1,
\end{equation}
To ensure orthogonality of such an eigenvector to those of the
longitudinal family, i.e. $(\eta^{(II)},\eta^{(I)})=0$, see
Eq.(\ref{scalar1}), amounts to imposing the conditions
\begin{equation}\label{fam2b}
\omega+(n-1)\nu=0,\quad\tau+\mu\,\frac{n-2}{2} =0,
\end{equation}
It is easy to check that these conditions in fact ensure that
\begin{equation}\label{cond1}
 \sum_{d}\eta^{(II)}_{(1d)}=-(n-1)(\nu-\tau),
\,\, \sum_{d}\eta^{(II)}_{(cd)}=\nu-\tau,\, \forall\,c\ne
1,\quad\mbox {so that} \quad \sum_{(cd)}\eta^{(II)}_{(cd)}= 0.
\end{equation}

Note that in the replica limit $n\to 0$ the relations
Eq.(\ref{fam2b}) imply $\omega=\nu$ and $\tau=\mu$. Then it is easy
to check that the equations Eq.(\ref{stab7aa},\ref{stab7bb}) for
$a=1$ are reduced to precisely the same system of two equations as
that in Eq.(\ref{long1},\ref{long2}), with the correspondence
$\alpha\to \nu,\quad \beta \to \tau$. Moreover, for any $a>1$ the
equations Eq.(\ref{stab7aa},\ref{stab7bb}) yield once more the same
pair of equations. We therefore conclude that the fluctuations
corresponding to this second family of eigenvectors are also unable
to induce instability of the replica-symmetric solution.

Finally, the third family of eigenvectors $\eta^{(III)}$ orthogonal
to the first two turns out to be characterised by a chosen pair of
nonequal replica indices $(a,b),\, a\ne b$. For the sake of
definiteness we can take $a=1,b=2$ when the corresponding
eigenvector components are given explicitly by:
\begin{equation}\label{fam3a}
\eta^{(III)}_{(aa)}=0,\,\forall\, a; \,\,
\eta^{(III)}_{(12)}=\eta^{(III)}_{(21)}=\xi,\,
\end{equation}
\[
\eta^{(III)}_{(1b)}=\eta^{(III)}_{(2b)}=\eta^{(III)}_{(b1)}=
\eta^{(III)}_{(b2)}=\psi,\,\,\forall\, b>2,\,
\]
\[\eta^{III}_{(ab)}=(1-\delta_{ab})\rho,\,\forall\, a>2
\,;\forall\,b>2
\]
where the parameters $\xi,\psi,\rho$ satisfy the constraints:
\begin{equation}\label{fam3b}
\xi+(n-2)\psi=0,\quad 2\psi+\rho\,(n-3)=0,
\end{equation}
It is easy to check that these conditions in fact ensure that
\begin{equation}\label{cond2}
 \sum_{d}\eta^{(III)}_{(ad)}=0, \forall\,a
\end{equation}
so that automatically $\sum_{(cd)}\eta^{(III)}_{(cd)}= 0.$ We then
find that the equations Eq.(\ref{stab7bb}) are satisfied
identically, whereas the equations Eq.(\ref{stab7aa}) are reduced to
the form
\begin{equation}\label{ATeig}
\left[(p_d-p_0)^2-\frac{1}{T^2}f''(D)\right]\eta^{(III)}_{(ab)}=
\Lambda^*\eta^{(III)}_{(ab)},\quad \forall \,a\ne b.
\end{equation}
As the replica symmetric solution in the $\mu-$dominated regime
implies $D=T/\mu,\, p_d-p_0=\mu/T$, see Eqs.(\ref{invsym},
\ref{qsym}), we infer from Eq.(\ref{ATeig}) that the stability
condition $\Lambda^*\ge 0$ indeed implies the inequality
Eq.(\ref{ATmu}).

Turning to the stability of $R-$dominated replica-symmetric
solution, we find that the eigen-equations Eq.(\ref{stab8}) are
reduced to
\begin{equation}\label{stab8aa}
p_0^2\,\sum_{(cd)}\eta_{(cd)}+p_0(p_d-p_0)\,\left[\sum_{d}\eta_{(ad)}+
\sum_{d}\eta_{(bd)}\right]+(p_d-p_0)^2\,\eta_{(ab)}
\end{equation}
\[
-\frac{1}{T^2} f''\left(R^2-q_0\right)\eta_{(ab)}
=\Lambda^*\,\eta_{(ab)},\quad a\ne b
\]
The subsequent analysis is completely analogous to one performed
above in the $\mu-$dominated regime. In particular, the relevant
eigenvalue again corresponds to the third family of eigenvectors
satisfying the constraint Eq.(\ref{cond2}). This fact, together with
the relation $p_d-p_0=1/(R^2-q_0)$ immediately yields the inequality
Eq.(\ref{ATspherical}) as the corresponding stability condition.

\subsection{Stability of the Parisi solution with one step of the replica
symmetry breaking.}

In the following we will show that in this situation there are
essentially two relevant eigenvalues, both in the $\mu-$dominated
and $R-$dominated regimes:

\begin{equation}
\Lambda^*_K= {1\over (q_d-q_1)^2} - {1\over T^2} f''(q_d-q_1)
\label{lambdaK}
\end{equation}
and
\begin{equation}
  \Lambda^*_0= {1\over (q_d-q_1 +m(q_1-q_0))^2} - {1\over T^2} f''(q_d-q_0)
\label{lambda0}\ .
\end{equation}
If both are positive, all other eigenvalues are positive and the
system is stable. In the replica symmetric case both eigenvalues
fall together. Let us in general note that if we know the solution
on the line $T=T_b(\mu)$, which is defined by the condition
$q_d=R^2$, we know the solution in the whole $R$-dominated regime,
since there the values $q_d\ ,q_1\ ,q_0\ {\rm and}\  m$ are
independent of $\mu$ (see the main text of the paper). Thus it is in
general enough to proof the stability in the $\mu$-dominated regime.

To this end, let us consider the function $R(q)=S(q)/T$, where
$S(q)$ was defined in Eq.(\ref{con1}) of the main text. Specified
for the case of 1RSB this function satisfies (cf.(\ref{statkRSB})):
\begin{equation}
    R(q_0)=0,\ \ \ \ \ R(q_1)=0,\ \ \ \ \ {\rm and}\ \ \ \ \ \int_{q_0}^{q_1} R(q)\ dq =0\ .
\label{R(q)}
\end{equation}
Moreover we know that $R(q)$ must change sign once between $q_0$ and
$q_1$. Let us calculate the positions of the corresponding two
extrema $q_e^{(1,2)}$ from $R'(q_e)=0$:
\begin{equation}
 {1\over (q_d-q_1+m(q_1-q_e))^2} -{1\over T^2} f''(q_d-q_e)=0\ .
 \label{R'q*}
\end{equation}
which yields
\begin{equation}
q_d-q_1+m(q_1-q_e)= {T\over \sqrt{f''(q_d-q_e)}}
   \label{line}
\end{equation}
for $q_e=q_e^{(1,2)}$. Since the right-hand side is convex according
to our assumption for short range correlations, we have
$q_d-q_1+m(q_1-q)>{T/\sqrt{f''(q_d-q)}}$ for
$q_e^{(1)}<q<q_e^{(2)}$, hence $R'(q)<0$ in that interval.  This
however means that for $q\notin(q_e^{(1)},q_e^{(2)})$ necesserily
$R'(q)>0$, in particular $R'(q_0)>0$ and $R'(q_1)>0$. (The
accidental case $R'(q_0)=0$ or $R'(q_1)=0$ happens on the AT-line).
Note however that $R'(q_0)$ is exactly equal to $\Lambda^*_0$ and
similarly $R'(q_1)=\Lambda^*_K$. This observation proves stability
everywhere in the $\mu-$ dominated phase. The stability condition
becomes marginal along the AT-line.

In the rest of this appendix we will sketch the derivation of the
fluctuation eigenvalues in both regimes, $\mu$-dominated  and
$R$-dominated. First let us define a Parisi-block matrix $P_{ab}=1$
inside a diagonal Parisi $m\times m$ block and zero outside. With
this definition the entries of the Parisi matrix $Q$ has the form
\begin{equation}
   q_{ab}=q_0 + (q_1-q_0) P_{ab} +(q_d-q_1)\delta_ {ab}\ .
 \label{Parisiq}
\end{equation}
The inverse matrix $Q^{-1}$ has the same form:
\begin{equation}
  \left(Q^{-1} \right )_{ab}=A_0 + (A_1-A_0) P_{ab} +(A_d-A_1)\delta_ {ab}\ .
 \label{ParisiA}
\end{equation}

The matrix  $P_{ab}$ can be used to introduce the following
shorthand notations for several types of averages of the eigenvector
components $\eta_{(ab)}$ over the indices inside a Parisi block:
\begin{equation}
   \eta_{a\bar b}={1\over m}(\eta \ P)_{ab},\ \ \ \eta_{\bar a\bar b}={1\over m^2}(P \eta
P)_{ab},\ \ \ \eta_{\overline{aa}}={1\over m}\sum_{c}\eta_{(cc)} \
P_{ca}\ .
 \label{averages}
\end{equation}

With these definitions the eigen-equations Eq.(\ref{unified}) can be
written as
\begin{eqnarray}
\Lambda^* \eta_{ab}&=& A_0^2\sum_{c,d}\eta_{(cd)}
+(A_1-A_0)^2m^2\eta_{\bar a \bar b}+
(A_d-A_1)^2\eta_{(ab)}\nonumber \\
&&+mA_0(A_1-A_0)\sum_{c}(\eta_{\bar a c} + \eta_{c \bar b})+
A_0(A_d-A_1)\sum_{c}(\eta_{(ac)}+\eta_{(cb)})\\
&& +(A_1-A_0)(A_d-A_1)m(\eta_{ \bar a b} + \eta_{a \bar b}) \nonumber \\
\nonumber &&+ {1\over T^2}f''(D_{ab})\delta
D_{ab}-\delta_{ab}\sum_{c} {1\over T^2}f''(D_{ac})\delta D_{ac}.
\label{fluctuationmu}
\end{eqnarray}
Here, explicitly
\begin{equation}
f''(D_{ab})=f''(q_d-q_0)+(f''(q_d-q_1)-f''(q_d-q_0))P_{ab}
 \label{f''Q}
\end{equation}
and
\begin{eqnarray}
&&2\sum_{c} \nonumber f''(D_{ac})\delta D_{ac}=
f''(q_d-q_0)\left(n\,
\eta_{(aa)}+\sum_{c}\eta_{(cc)}-2\sum_{c}\eta_{(ac)}\right)+
\\  &+& m\left[ f''(q_d-q_1)-f''(q_d-q_0)\right]
\left(\eta_{(aa)}+\eta_{\overline {aa}}-2\eta_{a \bar a}\right).
 \label{sumf''Q}
\end{eqnarray}

Now one observes that one can derive a closed system of three
equations for the following three combinations
$$\sum_{a} \eta_{(aa)},\ \ \ \ \sum_{a}\eta_{\bar a \bar a},\ \ \ \ \sum_{ab}\eta_{(ab)}\ .$$
This yields three eigenvalues in the limit $n\to 0$:
 \begin{eqnarray}
\Lambda_1^*&=&(A_d-A_1+m(A_1-A_0))^2 \nonumber\\
\Lambda_2^*&=&(A_d-A_1)^2 -{1\over T^2} f''(q_d-q_1) +{1-m\over T^2}(f''(q_d-q_1)-f''(q_d-q_0))\nonumber\\
 \Lambda_3^*&=&(A_d-A_1+m(A_1-A_0))^2 -{1\over T^2} f''(q_d-q_0)\ .
  \label{eigenv1-3}
\end{eqnarray}

As the next family of eigenvectors we use those satisfying the
constraint $\sum_{a} \eta_{(aa)}  = \sum_{a}\eta_{\bar a \bar a}=
\sum_{ab}\eta_{(ab)}\ \ =0$. One can show that constraints of this
sort ensure orthogonality of the new families of eigenvectors to the
old one (compare with the de-Almeida-Thouless analysis of the
previous section). Subsequently, one derives a system of equations
for
$$ \eta_{\overline{aa}},\ \ \ \ \eta_{\bar a \bar a},\ \ \ \ \sum_{b}\eta_{\bar a  b}\ .$$
In the limit $n\to 0$ that system yields the same eigenvalues
$\Lambda^*_1, \Lambda^*_2, \Lambda^*_3$.

Using now as the constraint the conditions $ \eta_{\overline{aa}} =
\eta_{\bar a \bar a}= \sum_{b}\eta_{\bar a b}\ \ =0$ simultaneously
for all $a$ we derive a single equation for the component
$\eta_{\bar a \bar b}$, with the indices $a$ and $b$ in different
Parisi blocks (we write $\bar a \neq \bar b$). The resulting
eigenvalue is again $\Lambda^*_3$.

In the next step we impose another constraint $\eta_{\bar a \bar
b}=0$ for all $a,b$ to obtain a set of equations for
$$\eta_{(aa)},\ \ \ \ \eta_{a \bar a},\ \ \ \
\sum_{b}\eta_{(ab)}\ .$$ This leads to another three eigenvalues
\begin{eqnarray}
\Lambda^*_4&=&(A_d-A_1)(A_d-A_1+m(A_1-A_0))  \nonumber\\
\Lambda^*_5&=&(A_d-A_1)^2 -{1\over T^2} f''(q_d-q_1) +{2-m\over 2T^2}(f''(q_d-q_1)-f''(q_d-q_0))\nonumber\\
 \Lambda^*_6&=&(A_d-A_1)(A_d-A_1+m(A_1-A_0))-{1\over T^2} f''(q_d-q_0)\ .
\label{eigenv4-6}
\end{eqnarray}

We then obtain a single equation for $\eta_{a \bar b}$, which leads
for $\bar a \neq \bar b$ to $\Lambda^*_6$ again. The remaining
equation for $\eta_{(ab)}$ yields for $\bar a \neq \bar b$ a new
eigenvalue
\begin{equation}
  \Lambda^*_7 = (A_d-A_1 )^2 -{1\over T^2} f''(q_d-q_0)
\label{lambda7}
\end{equation}
and for $\bar a = \bar b$ and $ a \neq b$
 \begin{equation}
 \Lambda^*_8 = (A_d-A_1)^2 -{1\over T^2} f''(q_d-q_1)\ .
 \label{lambda8}
\end{equation}
Assuming $f''(x)$ monotonously decreasing, $0<m<1$ and
$0<q_0<q_1<q_d$ the relevant eigenvalues are obviously
$\Lambda^*_2=\Lambda^*_0$ and $\Lambda^*_8 = \Lambda^*_K$. Here we
use that the eigenvalues of the matrix $Q$ are
\begin{equation}
 q_d-q_1,\ \ \   q_d-q_1 +m(q_1-q_0),\ \ \ q_d-q_1 +m(q_1-q_0)+nq_0
 \label{eigenvq}
\end{equation}
These are inverses of the eigenvalues of the matrix $Q^{-1}$:
\begin{equation}
 A_d-A_1,\ \ \   A_d-A_1+m(A_1-A_0),\ \ \ A_d-A_1 +m(A_1-A_0)+nA_0\ .
 \label{eigenvA}
\end{equation}

\bigskip
Turning now to the $R$-dominated regime, the eigenvalue equations
(\ref{stab8}) with the use of the short-hand notations introduced in
Eq.(\ref{DD}) can be written in the form:
\begin{eqnarray}
\Lambda^*\, \delta D_{ab}&=&  (A_0^2\sum_{c,d}\delta D_{c,d}
+(A_1-A_0)^2m^2\delta D_{\bar a \bar b} +(A_d-A_1)^2\delta
D_{ab}\nonumber \\ &+&mA_0(A_1-A_0)
\sum_{c}(\delta D_{\bar a  c} + \delta Q_{c \bar b}) +A_0(A_d-A_1) \sum_{c}(\delta D_{ac} + \delta D_{cb})\nonumber\\
&+& (A_1-A_0)(A_d-A_1)m(\delta D_{ \bar a  b} + \delta D_{a \bar b})
\nonumber  )(1-\delta_{ab})-{1\over T^2}f''(D_{ab}) \delta D_{ab}
 \label{fluctuationR}
\end{eqnarray}

We proceed in the same way as before. The equations for
$$ \sum_{a}\delta D_{\bar a \bar a},\ \ \ \ \sum_{ab}\delta D_{ab}$$
produce the quadratic equation $\left|\begin{array}{cc} A-\Lambda^*
& B \cr C & D-\Lambda^*\end{array}\right|=0$ with
\begin{eqnarray}
A&=&(m-1)(m(A_1-A_0)^2+2( A_1-A_0)(A_d-A_1)) + (A_d-A_1)^2-{1\over T^2}f''(q_d-q_1)\nonumber\\
B&=&(m-1)(nA_0^2 + 2A_0 (A_d-A_1+m(A_1-A_0)))\nonumber\\
C&=& -m(A_1-A_0)^2-2( A_1-A_0)(A_d-A_1)-{1\over T^2}f''(q_d-q_1)+{1\over T^2}f''(q_d-q_0)\nonumber\\
\nonumber D&=&n(n-1)A_0^2 +2(n-1)A_0(A_d-A_1+m(A_1-A_0))
+(A_d-A_1+m(A_1-A_0))^2\\ \nonumber && -{1\over T^2} f''(q_d-q_0)\,.
\end{eqnarray}

In the next step we obtain for
$$  \delta D_{\bar a \bar a},\ \ \ \ \sum_{b}\delta D_{\bar a b} $$ another quadratic equation
$  \left|\begin{array}{cc} A-\Lambda^* & B \cr C &
D-\Lambda^*\end{array} \right|=0$ with
\begin{eqnarray}
A&=&(m-1)(m(A_1-A_0)^2+2( A_1-A_0)(A_d-A_1))\ +\ (A_d-A_1)^2-{1\over T^2}f''(q_d-q_1)\nonumber\\
B&=&(m-1)(  2A_0 (A_d-A_1+m(A_1-A_0)))\nonumber\\
C&=& -m(A_1-A_0)^2-2( A_1-A_0)(A_d-A_1)-{1\over T^2}f''(q_d-q_1)+{1\over T^2}f''(q_d-q_0)\nonumber\\
\nonumber D&=&(n-2)A_0(A_d-A_1+m(A_1-A_0))
+(A_d-A_1+m(A_1-A_0))^2-{1\over T^2}f''(q_d-q_0)
\end{eqnarray}

Both equations fall together for $n\to 0$  and produce the same
eigenvalues $\Lambda^*_1,\Lambda^*_2$. One can show under the
conditions we have that both eigenvalues are positive provided
$\Lambda^*_0$ and $\Lambda^*_K$ are. Next one derives a single
equation for $\delta D_{\bar a  \bar b}$ for $\bar a \neq \bar b$
leading to the eigenvalue
\begin{equation}
\Lambda^*_3= (A_d-A_1+m(A_1-A_0))^2 -{1\over T^2} f''(q_d-q_0)
 \label{lambda3}
\end{equation}
Note that $\Lambda^*_3$ here is formally equal to $\Lambda^*_3$
found in the previous analysis of the $\mu-$dominated regime. On the
next lower level we obtain for
 $$  \delta D_{a a},\ \ \ \ \sum_{b}\delta D_{ab} $$
the eigenvalue equation $ \left|\begin{array}{cc} A-\Lambda^* & B
\cr C & D-\Lambda^*
\end{array}\right|=0$ with
\begin{eqnarray}
A&=&(m-2) ( A_1-A_0)(A_d-A_1))\ +\ (A_d-A_1)^2-{1\over T^2}f''(q_d-q_1)\nonumber\\
B&=&(m-2)  A_0 (A_d-A_K)  \nonumber\\
C&=&  -2( A_1-A_0)(A_d-A_1)-{1\over T^2}f''(q_d-q_1)+{1\over T^2}f''(q_d-q_0)\nonumber\\
D&=&  (A_d-A_1 )((n-2)A_0+m(A_1-A_0)) +(A_d-A_1 )^2-{1\over
T^2}f''(q_d-q_0)\nonumber
\end{eqnarray}
leading to $\Lambda^*_4,\Lambda^*_5$. Again both are positive if
$\Lambda^*_0$ and $\Lambda^*_K$ are. On the next lower levels we
again reproduce the previous eigenvalues $\Lambda^*_6$,
$\Lambda^*_7$ and $\Lambda^*_8$ of the $\mu-$dominated analysis. As
the result in both regimes are $\Lambda^*_0$ and $\Lambda^*_K$ the
relevant eigenvalues, which have to be positive for stability - and
they are indeed positive as we have demonstrated before.

\end{document}